\documentclass[%
    prd,      
    preprint, 
    showkeys, 
    superscriptaddress, 
    nofootinbib 
]{revtex4}

\usepackage{amsmath,amsfonts,amssymb}
\usepackage{booktabs} 
\usepackage{csquotes} 
\usepackage{enumitem}
\usepackage{graphicx}
\usepackage{tabularx}
\usepackage{multirow}
\graphicspath{{./figures/}}

\usepackage{color}
\definecolor{verdes}{cmyk}{0.92,0,0.59,0.4}  
\definecolor{verdec}{cmyk}{0.92,0,0.59,0.15} 

\usepackage{hyperref}
\hypersetup{
    colorlinks=true,
    linkcolor=verdec,
    urlcolor=verdec,
    citecolor=verdec
}

\newcommand{\beq}{\begin{equation}}
\newcommand{\eeq}{\end{equation}}
\newcommand{\bea}{\begin{eqnarray}}
\newcommand{\eea}{\end{eqnarray}}
\newcommand{\barr}{\begin{array}}
\newcommand{\earr}{\end{array}}
\newcommand{\bc}{\begin{center}}
\newcommand{\ec}{\end{center}}
\newcommand{\bit}{\begin{itemize}}
\newcommand{\eit}{\end{itemize}}
\newcommand{\ben}{\begin{enumerate}}
\newcommand{\een}{\end{enumerate}}

\newcommand{\gev}{{\;{\rm GeV}}}
\newcommand{\tev}{{\;{\rm TeV}}}

\newcommand{\trans}[2]{\text{PT}_{#1}^{(#2\text{-step})}}

\newcommand{\hc}{\ensuremath\text{H.c.}}

\newcommand{\nsfo}{N_{\rm\scriptscriptstyle SFO} }
\newcommand{\mh}{m_h}
\newcommand{\mch}{M_{H^\pm}}
\newcommand{\mhh}{M_H}
\newcommand{\ma}{M_A}

\newcommand{\tb}{t_\beta}
\newcommand{\cba}{c_{\beta-\alpha}}

\newcommand{\package}[1]{\textsc{\small #1}}

\begin{document}
\preprint{KIAS-P26038}

\title{\color{verdes} Strong First-Order Electroweak Phase Transitions and Gravitational Waves in the Normal Two-Higgs-Doublet Model: \\
A Comparative Study of the Four Yukawa Types and \\
Thermal Resummation Schemes 
}

\author{Jin-Hwan Cho}
\email{chof@nims.re.kr}
\affiliation{National Institute for Mathematical Sciences, Daejeon 34047, Republic of Korea} 

\author{Dongjoo Kim}
\email{dongjookim.phys@gmail.com}
\affiliation{National Institute for Mathematical Sciences, Daejeon 34047, Republic of Korea} 
\affiliation{Department of Physics, Konkuk University, Seoul 05029, Republic of Korea} 

\author{Jinheung Kim}
\email{jhkim1216@kias.re.kr}
\affiliation{School of Physics, Korea Institute for Advanced Study,  Seoul 02455, Republic of Korea} 

\author{Soojin Lee}
\email{soojin.lee@gapp.nthu.edu.tw}
\affiliation{Department of Physics, National Tsing Hua University, Hsinchu 30013, Taiwan} 
\affiliation{Center for Theory and Computation, National Tsing Hua University, Hsinchu 30013, Taiwan} 

\author{Jeonghyeon Song}
\email{jhsong@konkuk.ac.kr}
\affiliation{Department of Physics, Konkuk University, Seoul 05029, Republic of Korea} 

\begin{abstract}
We present a comprehensive global analysis of strong first-order electroweak phase transitions (SFOEWPTs) and their associated stochastic gravitational-wave (GW) backgrounds within the Normal Scenario of the $CP$-conserving Two-Higgs-Doublet Model (2HDM) with softly broken $Z_2$ symmetry, where the lighter $CP$-even scalar is identified as the observed $125~\text{GeV}$ Higgs boson. Across all four Yukawa structures (Type-I, II, X, and Y), we track the finite-temperature vacuum evolution, transition dynamics, and GW signatures. To quantify the theoretical uncertainty associated with thermal resummation, we perform a detailed comparison between the Parwani and Arnold--Espinosa prescriptions. While both schemes find that single-step paths overwhelmingly dominate successful transitions and consistently favor the Higgs alignment limit, the resulting SFOEWPT parameter space exhibits a pronounced scheme dependence. The Arnold--Espinosa prescription severely restricts the viable parameter space (with upper bounds on the heavy-scalar masses below approximately 800~GeV) and introduces an extreme parametric sensitivity that produces fragmented distributions and irregular voids in the heavy-scalar mass planes. In contrast, the more stable Parwani prescription allows heavy-scalar masses below $\sim 1.6~\text{TeV}$. We further identify highly restricted GW parameter regions capable of yielding a four-year LISA signal-to-noise ratio above 10, while demonstrating that the acoustic GW source is generically short-lived, leading to a substantial suppression of the predicted signal amplitude. Our results highlight the strong complementarity between future space-based GW observations and high-energy collider searches in probing the cosmological viability of the 2HDM.
\end{abstract}


\maketitle

\tableofcontents

\section{Introduction}

The origin of the observed baryon asymmetry of the Universe, characterized by the baryon-to-photon ratio $n_B/n_\gamma \approx 6 \times 10^{-10}$~\cite{Planck:2015fie}, remains a fundamental puzzle in particle physics and cosmology. Dynamically generating this asymmetry requires a strong first-order electroweak phase transition (SFOEWPT)~\cite{Huet:1994jb,Kajantie:1996mn} for the necessary departure from thermal equilibrium~\cite{Sakharov:1967dj}. An SFOEWPT can also source a stochastic gravitational-wave (GW) backgrounds~\cite{Weir:2017wfa,Caprini:2018mtu} potentially detectable by next-generation space-based interferometers such as the Laser Interferometer Space Antenna (LISA)~\cite{LISA:2017pwj,Cutting:2018tjt,Guo:2020grp,Schmitz:2020syl}. Within the Standard Model (SM), however, the $125\gev$ Higgs boson mass implies that the electroweak symmetry breaking occurs via a smooth crossover~\cite{Huet:1994jb,Kajantie:1996mn,Csikor:1998eu}. This tension strongly motivates extended Higgs sectors capable of facilitating an SFOEWPT~\cite{Bochkarev:1990fx,Dorsch:2013wja,Basler:2016obg,Fuyuto:2017ewj,Bernon:2017jgv,Kainulainen:2019kyp,Bittar:2025lcr}. 
Such sectors are particularly interesting in the context
of the EWPT, as they can support multi-step phase transitions in which the
Universe evolves through one or more intermediate metastable vacua before
settling into the electroweak minimum~\cite{Land:1992sm,Patel:2012pi,Inoue:2015pza,Blinov:2015sna,Bian:2017wfv,Chao:2017vrq,Ramsey-Musolf:2017tgh,Chala:2018opy,Morais:2019fnm,Fabian:2020hny,Aoki:2021oez,Goncalves:2021egx,Si:2024vrq}.

The Two-Higgs-Doublet Model (2HDM) provides a highly motivated framework for such extensions~\cite{Branco:2011iw}, introducing five physical scalar states: two CP-even ($h, H$), one CP-odd ($A$), and a pair of charged Higgs bosons ($H^\pm$). The dangerous flavor-changing neutral currents (FCNCs) at tree level are prevented by a softly broken discrete $Z_2$ symmetry~\cite{Glashow:1976nt,Paschos:1976ay}, yielding four distinct Yukawa realizations (Type-I, II, X, and Y). 
While Type-I remains comparatively weakly constrained, Type-II and Type-Y are subject to severe bounds, most notably a strict lower limit on the charged Higgs boson mass ($M_{H^\pm} \gtrsim 580\gev$) driven by radiative $B$-meson decays ($b \to s \gamma$)~\cite{Haller:2018nnx,Misiak:2020vlo,Biekotter:2024ykp,Biekotter:2025fjx}. On the other hand, Type-X experiences a milder yet significant restriction due to enhanced leptonic couplings that yield characteristic multi-$\tau$ collider signatures~\cite{ATLAS:2017eiz,Kanemura:2011kx,Abe:2015oca,CMS:2018rmh,ATLAS:2020zms,Kim:2023lxc}.
 These constraints tightly restrict the so-called Inverted Scenario—where the observed $125\gev$ state is identified as the heavier neutral scalar $H$—to be viable almost exclusively within Type-I.
 
Our previous study on the SFOEWPT and GW signals focused exclusively on the Inverted Scenario of the 2HDM, which is restricted to Type-I alone, leaving no room for a comparative study across different Yukawa structures~\cite{Lee:2025hgb}. In that work, from the physically viable parameter configurations satisfying all theoretical and collider requirements, we identified those supporting an SFOEWPT (referred to as the SFOEWPT parameter points) and those yielding a detectable GW signal (referred to as the GW parameter points).  Crucially, these SFOEWPT parameter points consistently favor light beyond-the-Standard-Model (BSM) scalars regardless of whether the transition completes in one or two steps. On the contrary, the resulting volume of the GW parameter points was heavily suppressed, yielding only 114 GW parameter points  out of the $2.36 \times 10^6$ physically viable points.

In this work, we substantially extend the cosmological investigation of the
2HDM by focusing on the Normal Scenario, in which the lighter CP-even scalar
$h$ is identified with the observed $125\gev$ Higgs boson. We present four
major advances beyond the existing literature. 

First, we perform a comprehensive comparative analysis of all four Yukawa
structures (Type-I, II, X, and Y).\footnote{We note that
Ref.~\cite{Aoki:2021oez} demonstrated the theoretical existence of multi-step
EWPTs across all four 2HDM types using specific benchmark configurations
(e.g., $M_H=M_{H^\pm}$) in the exact Higgs alignment limit, without imposing
current experimental constraints.} A particular emphasis is placed on the
realization of multi-step SFOEWPTs in the Normal Scenario. 
While our previous study found that the Inverted Scenario accommodates two-step
transitions at a sizable rate, with a frequency of approximately $80\%$ relative
to the one-step cases, the viability of multi-step EWPTs in the Normal Scenario has not yet been
systematically investigated across all four Yukawa structures.
We fill this gap by performing the first global
study of one-step and multi-step SFOEWPT histories in the Normal Scenario.

Crucially, the structural characteristics of these valid SFOEWPT parameter 
spaces reveal a striking, universal feature across all four Yukawa structures: 
the requirement of an SFOEWPT entirely truncates the conventional decoupling 
limit, which would otherwise permit arbitrarily heavy and nearly degenerate 
scalar states. At the same time, the lower bounds on the heavy-scalar masses, 
the allowed deviations from Higgs alignment, and the viable ranges of $\tan\beta$ 
exhibit a strong dependence on the specific Yukawa structure. These differences 
have important, model-dependent implications for future collider searches.

Second, we investigate the theoretical uncertainty associated with the thermal
resummation of daisy diagrams in the finite-temperature effective potential. We
compare two prescriptions widely used in the literature: the Parwani
method~\cite{Parwani:1991gq} and the Arnold--Espinosa (AE)
method~\cite{Arnold:1992rz}. Although many phenomenological analyses,
including our prior work, have adopted the AE scheme~\cite{Biekotter:2025fjx,
Braathen:2025svl,Lee:2025hgb,Anisha:2025zbc,Bittar:2025lcr,
Branchina:2025jou}, recent field-theoretic developments suggest that the
Parwani scheme is more closely aligned with self-consistent partial-dressing
formalisms~\cite{Boyd:1993tz,Curtin:2016urg,Curtin:2022ovx,Bahl:2024ykv,
Bittar:2025lcr}. By comparing the Parwani and AE schemes across both key model
parameters and macroscopic phase-transition parameters, such as the transition
strength $\alpha_{\rm GW}$ and the inverse duration $\beta_{\rm GW}/H_*$, we
provide a systematic assessment of resummation-induced theoretical
uncertainties. We further show that the AE scheme induces a severe sensitivity
of the percolation order parameter, $\xi_p$, to small variations in the heavy
BSM scalar masses, exposing a significant limitation of this prescription.
Such a side-by-side comparison of the two thermal-resummation schemes has
been absent from the literature.

The third major contribution is a detailed investigation of the plasma
hydrodynamics underlying the GW parameter points. In the 2HDM, the dominant
contribution to the GW spectrum arises from acoustic sound waves, so the
observable signal is largely controlled by macroscopic fluid dynamics. Rather
than focusing only on the final GW signal strength, we examine how the
transition duration, acoustic kinetic-energy fraction, degree of supercooling,
and sound-wave lifetime vary with the percolation order parameter $\xi_p$. This
allows us to identify the hydrodynamic mechanisms that give rise to highly
observable GW signals.

Finally, we establish a robust and comprehensive high-statistics global scanning methodology to ensure unbiased, structurally sound conclusions regarding the SFOEWPT and GW landscapes across the full parameter space of the Normal 2HDM. By subjecting a vast ensemble of $10^6$ physically viable parameter configurations per Yukawa type to rigorous finite-temperature calculations using \package{BSMPT} v3.1.8~\cite{Basler:2018cwe,Basler:2020nrq,Basler:2024aaf} under both resummation frameworks, our analysis handles a total of 8 million distinct thermodynamic histories. This framework represents one of the largest phase-transition landscape scans executed within the 2HDM to date, providing a statistically definitive baseline for future phenomenological studies.

The remainder of this paper is organized as follows.
Section~\ref{sec-review} briefly reviews the 2HDM framework, the finite-temperature
one-loop effective potential, the macroscopic phase-transition parameters, and
the stochastic GW observables. In Sec.~\ref{sec-scan-method}, we describe our
global scanning procedure and provide a statistical overview of the resulting
dataset across all four Yukawa types and both resummation schemes. This includes
the total numbers of the SFOEWPT and GW parameter points, the maximal $\xi_p$, and the peak SNR.
Section~\ref{sec:PvsAE} assesses the scheme dependence of the SFOEWPT parameter
space by comparing the Parwani and AE prescriptions. In
Sec.~\ref{sec:characteristics:SFOEWPT:Parwani}, we compare the characteristics
of the SFOEWPT parameter space across the four Yukawa types within the more
stable Parwani prescription. Section~\ref{sec:characteristics:GW:Parwani}
examines the plasma hydrodynamics and phenomenological features underlying the
successful GW configurations. Finally, Sec.~\ref{sec-conclusion} summarizes our
main findings and conclusions. The supplementary selection criteria applied to
the raw \package{BSMPT} outputs to ensure cosmological viability are detailed in
Appendix~\ref{appendix-BSMPT-selection}.

\section{Review of SFOEWPT in the 2HDM}
\label{sec-review}

\subsection{Brief review of the 2HDM}

The 2HDM extends the SM by introducing two complex $SU(2)_L$ scalar doublets, $\Phi_1$ and $\Phi_2$, both carrying hypercharge $Y=1/2$ under the convention $Q = T_3 + Y$~\cite{Branco:2011iw}.
These fields are expressed as:
\begin{equation} 
\label{eq:phi:fields}
\Phi_i = \begin{pmatrix} w_i^+ \\ \dfrac{v_i + \rho_i + i \eta_i}{\sqrt{2}} \end{pmatrix}, \quad (i=1,2),
\end{equation}
where $v_1$ and $v_2$ are the vacuum expectation values (VEVs) of $\Phi_1$ and $\Phi_2$, respectively.
Their ratio defines a key model parameter, $\tb = v_2/v_1$, where we use $s_x = \sin x$, $c_x = \cos x$, and $t_x = \tan x$ for notational simplicity.
The combined VEV, $v = \sqrt{v_1^2+v_2^2} \approx 246\gev$, induces spontaneous electroweak symmetry breaking.

To avoid tree-level FCNCs, a discrete $Z_2$ symmetry is imposed, under which $\Phi_1 \to \Phi_1$ and $\Phi_2 \to -\Phi_2$~\cite{Glashow:1976nt,Paschos:1976ay}.
The scalar potential, respecting \textit{CP} invariance and soft breaking of this $Z_2$ symmetry, is given by:
\begin{equation}
\label{eq:VPhi}
\begin{split}
V_\Phi (\Phi_1,\Phi_2) &=  m^2_{11} \Phi^\dagger_1 \Phi_1 + m^2_{22} \Phi^\dagger_2 \Phi_2 - m^2_{12} (\Phi^\dagger_1 \Phi_2 + \hc) \\[3pt]
&\quad + \tfrac{1}{2}\lambda_1 (\Phi^\dagger_1 \Phi_1)^2 + \tfrac{1}{2}\lambda_2 (\Phi^\dagger_2 \Phi_2)^2  + \lambda_3 (\Phi^\dagger_1 \Phi_1) (\Phi^\dagger_2 \Phi_2)  \\[3pt]
&\quad + \lambda_4 (\Phi^\dagger_1 \Phi_2)(\Phi^\dagger_2 \Phi_1) + \tfrac{1}{2} \lambda_5 \bigl[ (\Phi^\dagger_1 \Phi_2)^2 + \hc \bigr],
\end{split}
\end{equation}
where the $m_{12}^2$ term softly breaks the $Z_2$ symmetry.
The mixing angles $\alpha$ and $\beta$ parameterize the rotations between the interaction and mass eigenstates, yielding the physical scalar spectrum consisting of the \textit{CP}-even states ($h,H$), the \textit{CP}-odd state ($A$), and the charged scalars ($H^\pm$).
For their precise definitions, we refer to our previous study~\cite{Lee:2025hgb}.

The SM Higgs boson is  a linear superposition of the \textit{CP}-even eigenstates $h$ and $H$:
\beq
\label{eq-hsm-def} 
h_\text{SM} = s_{\beta-\alpha} h + \cba H.
\eeq
We focus on the `Normal Scenario', where the lighter scalar $h$ is identified as the observed 125 GeV Higgs boson:
\beq
\label{eq-model}
\text{Normal Scenario:}\quad \mh=125\gev.
\eeq
In this Normal Scenario, the gauge coupling modifier $\kappa_V$ ($V=W^\pm, Z$) becomes
\beq
\label{eq-kappaV}
\kappa_V = s_{\beta-\alpha}.
\eeq
The exact Higgs alignment limit corresponds to $s_{\beta-\alpha}= \pm 1$, for which the gauge and Yukawa couplings of the 125~GeV Higgs boson coincide with those of the SM Higgs boson, up to an overall
sign.
This alignment limit is strongly motivated by experimental constraints from the LHC, which confirm the SM-like behavior of the observed $125\gev$ scalar~\cite{ATLAS:2012yve,CMS:2012qbp}. Theoretically, it is equally compelling because such an alignment can arise naturally from global symmetries embedded within the scalar potential~\cite{BhupalDev:2014bir}.

In this framework, the phenomenology is determined by the following six independent parameters:
\begin{equation}
\label{eq-model-parameters}
\big\{ \tb,\; s_{\beta-\alpha},\; m_{12}^2,\;  \mhh,\;  \ma,\; \mch \big\},
\end{equation}
which uniquely determine the scalar quartic couplings $\lambda_i$ ($i=1,\dots,5$). We adopt the analytic relations given in Ref.~\cite{Lee:2025hgb}.

The 2HDM admits four Yukawa realizations (Type-I, II, X, and Y), distinguished by the $Z_2$ charge assignments of the right-handed fermions.
For the Yukawa Lagrangian parametrized as:
\begin{equation}
\begin{split}
\mathcal{L}_\text{Yuk} &=  - \sum_{f=u,d,\tau} \Bigl(
\frac{m_f}{v} \kappa_f \bar{f} f h + \frac{m_f}{v} \xi^H_f \bar{f} f H - i \frac{m_f}{v} \xi^A_f \bar{f} \gamma_5 f A 
\Bigr) \\
&\quad - \Bigl\{
\dfrac{\sqrt{2}V_{ud}}{v} H^+ \overline{u} (m_u \xi^A_u \text{P}_L +  m_d \xi^A_d \text{P}_R)d
+ \dfrac{\sqrt{2} m_\tau}{v} H^+ \xi^A_\tau \overline{\nu}_L\tau_R  + \hc
\Bigr\},
\end{split}
\end{equation}
the Yukawa coupling modifiers are presented in \autoref{tab-Yukawa}.

\begin{table}[h]
\centering
\renewcommand{\arraystretch}{1.3} 
\setlength{\tabcolsep}{12pt}
\begin{tabular}{l|rrr|ccc|rrr}
\hline \hline
 & $\kappa_u$ & $\kappa_d$ & $\kappa_\tau$ & $\xi_u^H$ & $\xi_d^H$ & $\xi_\tau^H$ & $\xi_u^A$ & $\xi_d^A$ & $\xi_\tau^A$ \\
\hline
Type I & $\frac{c_\alpha}{s_\beta}$ & $\frac{c_\alpha}{s_\beta}$ & $\frac{c_\alpha}{s_\beta}$ & $\frac{s_\alpha}{s_\beta}$ & $\frac{s_\alpha}{s_\beta}$ & $\frac{s_\alpha}{s_\beta}$ & $\frac{1}{\tb}$ & $-\frac{1}{\tb}$ & $-\frac{1}{\tb}$ \\
Type II & $\frac{c_\alpha}{s_\beta}$ & $-\frac{s_\alpha}{c_\beta}$ & $-\frac{s_\alpha}{c_\beta}$ & $\frac{s_\alpha}{s_\beta}$ & $\frac{c_\alpha}{c_\beta}$ & $\frac{c_\alpha}{c_\beta}$ & $\frac{1}{\tb}$ & $\tb$ & $\tb$ \\
Type X & $\frac{c_\alpha}{s_\beta}$ & $\frac{c_\alpha}{s_\beta}$ & $-\frac{s_\alpha}{c_\beta}$ & $\frac{s_\alpha}{s_\beta}$ & $\frac{s_\alpha}{s_\beta}$ & $\frac{c_\alpha}{c_\beta}$ & $\frac{1}{\tb}$ & $-\frac{1}{\tb}$ & $\tb$ \\
Type Y & $\frac{c_\alpha}{s_\beta}$ & $-\frac{s_\alpha}{c_\beta}$ & $\frac{c_\alpha}{s_\beta}$ & $\frac{s_\alpha}{s_\beta}$ & $\frac{c_\alpha}{c_\beta}$ & $\frac{s_\alpha}{s_\beta}$ & $\frac{1}{\tb}$ & $\tb$ & $-\frac{1}{\tb}$ \\
\hline \hline
\end{tabular}
\caption{Yukawa coupling modifiers in four different 2HDM Types.}
\label{tab-Yukawa}
\end{table}

\subsection{One-Loop Effective Potential at Finite Temperature}

In the early Universe, the dynamics of the electroweak phase transition is governed by the finite-temperature effective potential for spatially homogeneous classical scalar field configurations $\Phi_{1,2}^c$~\cite{Espinosa:1992kf,Herring:2024pqa,Chakrabortty:2024wto,Masina:2025pnp}. We parameterize these configurations in terms of three homogeneous real background fields, $w_1$, $w_2$, and $w_3$: 
\begin{equation}
\label{eq-classical-field}
\Phi_1^c = \frac{1}{\sqrt{2}}
\begin{pmatrix}
0 \\ w_1
\end{pmatrix},
\qquad
\Phi_2^c = \frac{1}{\sqrt{2}}
\begin{pmatrix}
0 \\ w_2 + i w_3
\end{pmatrix},
\end{equation}
where $w_3$ is retained to account for possible transient \textit{CP} violation during thermal evolution, despite the \textit{CP}-conserving nature of the zero-temperature vacuum~\cite{Ferreira:2015pfi,Basler:2016obg,Si:2024vrq}.

The full one-loop effective potential at temperature $T$ is given by 
\begin{equation} \label{eq-Veff} 
V_{\text{eff}}(\vec{w}, T) = V_{\text{tree}}(\vec{w}) + V_{\text{CW}}(\vec{w}) + V_{\text{CT}}(\vec{w}) + V_T(\vec{w}, T), 
\end{equation} 
where $\vec{w}=(w_1,w_2,w_3)$. The explicit forms of the tree-level potential $V_{\text{tree}}$, the Coleman--Weinberg potential $V_{\text{CW}}$, the counterterm potential $V_{\text{CT}}$, and the thermal corrections $V_T$ follow our previous work~\cite{Lee:2025hgb} and are not repeated here.

A major difference from our previous analysis lies in the treatment of thermal
daisy effects. These effects must be resummed to control the breakdown of the
high-temperature perturbative expansion caused by infrared divergences in the
bosonic sector. Two standard prescriptions are widely used for this purpose:
the Parwani method~\cite{Parwani:1991gq} and the AE
method~\cite{Arnold:1992rz}.

In the Parwani scheme, thermal screening effects are incorporated by replacing the field-dependent tree-level mass eigenvalues $m^2(\vec{w})$ with the thermally corrected Debye masses, 
\begin{equation} 
M^2(\vec{w},T)=m^2(\vec{w})+\Pi(T), 
\end{equation}
throughout both the zero-temperature and finite-temperature loop functions. Here, $\Pi(T)$ denotes the leading-order thermal self-energy corrections.

In contrast, the AE method performs a selective resummation of only the
Matsubara zero modes~\cite{Espinosa:1992gq}, which are responsible for the
infrared divergence of the bosonic thermal potential. Specifically, the
field-dependent masses squared of the $n=0$ modes are dressed by the
leading-order thermal self-energies, $\Pi(T)$, while the non-zero Matsubara
modes ($n\neq0$) are left unchanged. Consequently, only the field-dependent
cubic terms associated with the zero modes in the high-temperature expansion of
$V_T$ are modified, with the bare masses replaced by their thermally screened
counterparts.

Our previous study employed only the AE prescription, following much of the
phenomenological literature. The AE scheme has often been favored because the
global mass replacement in the Parwani method is regarded as a potential source
of overestimating the transition strength~\cite{Biekotter:2025fjx,Braathen:2025svl,Lee:2025hgb,Anisha:2025zbc,Bittar:2025lcr,Branchina:2025jou}. However, recent comparisons with advanced partial-dressing
formalisms~\cite{Boyd:1993tz,Curtin:2016urg,Curtin:2022ovx,Bahl:2024ykv}
suggest that, at least in the 2HDM, the Parwani prescription yields vacuum
profiles and phase-transition parameters closer to those obtained from
partial-dressing approaches~\cite{Bittar:2025lcr}. This motivates a systematic
assessment of the theoretical uncertainty associated with the choice of
daisy-resummation scheme.

For the purpose of estimating the theoretical uncertainty associated with daisy
resummation, we restrict our comparison to the Parwani and AE prescriptions and
do not implement partial dressing. While partial dressing provides a more
sophisticated treatment of thermal effects, it requires the iterative solution
of temperature-dependent non-linear gap equations for every parameter point
across the full thermal history. Given the scale of the present analysis, such
an approach is computationally prohibitive.

This scheme dependence can be analyzed using public codes.
\package{CosmoTransitions}~\cite{Wainwright:2011kj},
\package{PT2GWFinder}~\cite{Brdar:2025gyo}, and
\package{ELENA}~\cite{Costa:2025pew} employ the Parwani prescription by default,
whereas \package{BSMPT}~\cite{Basler:2018cwe,Basler:2020nrq,Basler:2024aaf} and
\package{PhaseTracer2}~\cite{Athron:2020sbe,Athron:2024xrh} adopt the AE prescription
as the default while providing a built-in option to switch to the Parwani scheme.
In this work, we use \package{BSMPT} v3.1.8.


\subsection{Characterizing SFOEWPTs and GW Signals}
\label{subsec:vcuum:GW}

Having established the finite-temperature effective potential, we now review the
quantities that characterize the vacuum evolution and the resulting stochastic GW
signals. The physical realization of the phase
transition proceeds via thermal tunneling, characterized by the nucleation of
bubbles of the true vacuum. 
This occurs at the nucleation temperature $T_n$, defined by the condition that
the expected number of critical bubbles nucleated within a Hubble volume
becomes of order unity, which is typically approximated by a bounce-action
criterion~\cite{Coleman:1977py,Callan:1977pt,Linde:1981zj,Bardsley:2021lmq,Athron:2022mmm}.
Following nucleation, the bubbles expand and merge. At the percolation temperature
$T_p$, the broken phase occupies roughly 29\% of the Universe. Thereafter, the false
vacuum continues to disappear as the transition proceeds toward completion.
Finally, the vacuum evolves to the zero-temperature global minimum, which in the
\textit{CP}-conserving limit corresponds to the standard VEVs of
$v_1=v\cos\beta$ and $v_2=v\sin\beta$.

In an extended Higgs sector, the electroweak phase transition may proceed
through multiple steps~\cite{Land:1992sm,Patel:2012pi,Inoue:2015pza,
Blinov:2015sna,Ramsey-Musolf:2017tgh,Chala:2018opy,Morais:2019fnm,
Fabian:2020hny}. To label the
individual transitions within an $n$-step EWPT, we introduce the notation
$\trans{i}{n}$:
\[
\trans{i}{n}: \thickspace \text{the $i$-th stage in an $n$-step SFOEWPT.}
\]
For a given transition step, we quantify the transition
strength by the dynamical order parameter at the percolation temperature,
\beq
\xi_p \equiv \frac{\Delta v(T_p)}{T_p},
\eeq
where $\Delta v(T_p)$ denotes the magnitude of the change in the VEVs between
the two successive phases. For a one-step EWPT, $\Delta v(T_p)=|\vec{w}(T_p)|$
corresponds to the field-excursion distance from the symmetric phase. We define
an SFOEWPT by the criterion\footnote{In our previous work, we adopted the SFOEWPT criterion based on the order parameter evaluated at the critical temperature, $\xi_c$.}
\beq
\label{eq-SFOEWPT-condition}
\text{SFOEWPT:}\quad \xi_p > 1.
\eeq
This threshold is adopted to ensure sufficient suppression of
sphaleron-induced washout of any baryon asymmetry generated during the
electroweak phase transition.

Beyond determining the thermal history of the Universe, an SFOEWPT can source a
stochastic GW background whose spectral properties are primarily governed by three
macroscopic parameters evaluated at the transition temperature $T_*$, which we
identify with the percolation temperature $T_p$: the phase transition strength
$\alpha_{\text{GW}}$, the inverse duration normalized to the Hubble rate
$\beta_{\text{GW}}/H_*$, and the bubble wall velocity $v_\text{w}$~\cite{Caprini:2019egz,Auclair:2022lcg,Basler:2024aaf}. Throughout this work, we
use the subscript ``GW'' for $\alpha$ and $\beta$ to distinguish these thermal
parameters from the 2HDM mixing angles.

The transition strength $\alpha_{\text{GW}}$ is defined as the ratio of the latent
heat $\epsilon(T_*)$ released during the transition to the radiation energy density
of the Universe $\rho_{\text{rad}}(T_*)$~\cite{Hindmarsh:2015qta,Hindmarsh:2017gnf}:
\begin{equation}
\label{eq-alphaGW}
    \alpha_{\text{GW}}(T_*) =
    \frac{\epsilon(T_*)}{\rho_{\text{rad}}(T_*)}
    =
    \frac{1}{\rho_{\text{rad}}(T_*)}
    \left[
    \Delta V(T)
    -
    \frac{T}{4}
    \frac{\partial \Delta V(T)}{\partial T}
    \right]_{T=T_*},
\end{equation}
where $\Delta V(T)\equiv V(\vec{w}_{\rm false})-V(\vec{w}_{\rm true})>0$
denotes the difference in the effective potential between the false and true
phases. The radiation energy density is
$\rho_{\text{rad}}(T_*)=g_*\pi^2 T_*^4/30$, where $g_*$ is the effective number of
relativistic degrees of freedom in the plasma at $T_*$. In the 2HDM, we adopt the
benchmark value $g_*=110.75$, which serves as a practical upper bound assuming
that all additional scalar degrees of freedom are fully relativistic at the
transition temperature. The definition of $\alpha_{\text{GW}}$ in
\autoref{eq-alphaGW} implies the thermodynamic consistency condition~\cite{Quiros:1999jp}
\begin{equation}
\label{eq-alphaGW-positive}
    \alpha_\text{GW} > 0.
\end{equation}

The second key parameter, $\beta_{\text{GW}}/H_*$, characterizes the inverse duration
of the phase transition in units of the Hubble time at $T_*$. It is defined by the
temperature derivative of the three-dimensional, $O(3)$-symmetric Euclidean bounce
action $S_3$ evaluated at the transition temperature~\cite{Ellis:2018mja}:
\begin{equation}
 \frac{\beta_{\text{GW}}}{H_*}
=
 T_*
 \left.
 \frac{d}{dT}
 \left(
 \frac{S_3(T)}{T}
 \right)
 \right|_{T=T_*},
\end{equation}
where $H_*$ is the Hubble parameter at $T_*$, including the contribution from the vacuum
energy density:
\[
H_* = \frac{1}{\sqrt{3}\,M_{\rm pl}}\sqrt{\frac{g_* \pi^2}{30}T_*^4 + \Delta V (T_*)}.
\]
 
A smaller value of
$\beta_{\text{GW}}/H_*$ denotes a slower transition with a more prolonged transition
duration. When this parameter drops below unity ($\beta_{\text{GW}}/H_* < 1$), 
the phase transition duration exceeds the Hubble time.
Since such ultra-slow 
transitions can generate super-horizon gravitational wavelengths, directly violating 
causality bounds on the stochastic signal amplitude~\cite{Caprini:2006jb,Athron:2023xlk},
\package{BSMPT} enforces a strict lower bound of $\beta_{\text{GW}}/H_* \ge 1$ 
for all viable GW evaluations: see Appendix \ref{appendix-BSMPT-selection}.

The final parameter is the bubble wall velocity $v_\text{w}$ after
nucleation, measured in the rest frame of the cosmological plasma. This parameter
is critical to the resulting percolation temperature and GW spectrum~\cite{Ai:2021kak,Dorsch:2021nje,Jiang:2022btc}.
While recent developments have provided new insights into the microphysics of
bubble expansion~\cite{Lewicki:2021pgr,Laurent:2022jrs}, predicting $v_\text{w}$
from first principles remains challenging due to its sensitivity to
model-dependent friction effects.
Throughout this work, we adopt a benchmark value\footnote{We note a numerical inconsistency in the public release of \package{BSMPT} v3.1.8
associated with the treatment of the bubble wall velocity. Although
\package{BSMPT} allows the user to specify the wall velocity, the initial
phase-tracking routine evaluating the percolation and completion
temperatures uses a fixed value $v_{\text{w}}=0.95$, rather than the user-specified
input. Since the time-dependent false vacuum fraction depends on the cube of the
wall velocity within the exponential suppression exponent,
$h(t)=\exp[-v_{\text{w}}^3\mathcal{F}_{\text{therm}}]$~\cite{Guth:1982pn,Turner:1992tz,Caprini:2019egz}, where
$\mathcal{F}_{\text{therm}}$ isolates the integrated thermodynamic dependence,
this can misclassify the phase transition history. We manually modified the source
code so that the phase-tracking routines consistently inherit the user-defined
bubble wall velocity. The required source modifications are documented in our
public contribution at
\url{https://github.com/BSMPT/BSMPT/issues/296\#issuecomment-4417065075}. The
downstream calculations in the internal \texttt{CalcGW} mode correctly incorporate
the user-specified $v_{\text{w}}$.}
of $v_\text{w}=0.6$.
This is a standard approximation for non-runaway transitions when the initial
acceleration phase is negligible compared with the overall bubble lifetime~\cite{Megevand:2009gh,Lee:2025hgb}. The chosen value lies slightly above the
speed of sound in a relativistic plasma, $c_s=1/\sqrt{3}$.

The total stochastic GW power spectrum is modeled
as the sum of three primary contributions~\cite{Caprini:2015zlo,Hindmarsh:2017gnf}:
\begin{equation}
    h^2 \Omega_{\text{GW}}(f)
    =
    h^2 \Omega_{\text{sw}}(f)
    +
    h^2 \Omega_{\text{col}}(f)
    +
    h^2 \Omega_{\text{turb}}(f),
\end{equation}
which arise from acoustic sound waves in the plasma bulk,
bubble-wall collisions, and magnetohydrodynamic turbulence, respectively. 
Here $h\simeq 0.674$ is the present-day dimensionless Hubble parameter~\cite{Planck:2018vyg}. 
For
SFOEWPTs with non-runaway bubble expansion, the sound-wave contribution is
generally the dominant component of the GW spectrum~\cite{Biekotter:2022kgf,Ramsey-Musolf:2024zex}.
We therefore focus our discussion on the acoustic GW spectrum, while our numerical
results include all three contributions.

The sound-wave GW amplitude is written as~\cite{Caprini:2024hue}
\begin{equation}
    \label{eq:Omega-dbpl}
    h^2 \Omega_{\text{sw}}(f)
    =
    h^2 \Omega_\text{int}
    S_{\text{sw}}(f)
    \Upsilon(H_* \tau_{\text{sw}})
    =
    h^2 \Omega_2
    S_{2}(f)
    \Upsilon(H_* \tau_{\text{sw}}),
\end{equation}
where $\Omega_\text{int}$ represents the integrated GW amplitude,
\begin{equation}
\label{eq-Omega}
h^2 \Omega_\text{int}
\simeq
5.284 \times 10^{-6}
\left( \frac{100}{g_*} \right)^{1/3}
v_\text{w}
\frac{H_*}{\beta_\text{GW}}
K_\text{sw}^2.
\end{equation}
In this expression, $K_\text{sw}$ denotes the global kinetic energy fraction of the 
total energy density of the Universe converted into acoustic waves:
\begin{equation}
K_\text{sw}
=
0.6 \, \frac{\kappa_\text{sw}\alpha_\text{GW}}{1+\alpha_\text{GW}},
\end{equation}
where  the numerical 
coefficient $0.6$ accounts for the efficiency reduction of kinetic energy production 
in the bulk fluid relative to the single-bubble configuration~\cite{Jinno:2022mie},
and $\kappa_{\text{sw}}$ is the sound-wave efficiency factor. 
An accurate determination of $\kappa_{\text{sw}}$ is therefore essential for reliable 
GW spectrum predictions. While earlier analyses frequently relied on auxiliary 
fitting functions ($\kappa_{A,B,C,D}$)~\cite{Basler:2024aaf}, recent versions of 
\package{BSMPT} (v3.1.0 and later) implement a model-independent energy-budget 
method~\cite{Giese:2020znk}. This numerical refinement significantly improves the treatment of macroscopic 
plasma hydrodynamics, particularly near the sonic boundary $v_{\text{w}}\simeq c_s$, 
where the transitions between the deflagration, hybrid, and detonation regimes are 
highly sensitive.

For the spectral shape function $S_{\text{sw}}(f)$, earlier studies often used a
single-peak broken power law~\cite{Guo:2020grp,Basler:2024aaf}. However, such
a description cannot simultaneously capture the large-scale bubble separation and
the finite thickness of the sound shell. We therefore model the acoustic GW
spectrum using a double broken power law~\cite{Hindmarsh:2017gnf,Hindmarsh:2019phv,Jinno:2022mie,Caprini:2024hue}, for which
\begin{equation}
    \label{eq:dbpl_sw_specific}
    S_{\text{sw}}(f)
    =
    N
    \left( \frac{f}{f_1} \right)^3
    \left[
    1+\left( \frac{f}{f_1} \right)^2
    \right]^{-1}
    \left[
    1+\left( \frac{f}{f_2} \right)^4
    \right]^{-1}.
\end{equation}
The normalization constant $N$ is determined by imposing
$S_{\text{sw}}(f_2)=1$~\cite{Caprini:2024hue}. The two characteristic frequency breaks,
$f_1$ and $f_2$, are governed by the mean bubble size $R_*$ and the finite thickness of the
fluid shell:\footnote{%
To obtain these expressions, we substitute the standard estimate for the mean
bubble radius at percolation,
$H_*R_*=\max(v_{\rm w},c_s)(8\pi)^{1/3}(H_*/\beta_{\rm GW})$,
into the frequency formulas of Ref.~\cite{Caprini:2024hue}. The corresponding
full expression  is
$
R_* =
\left[
T_*^3
\int_{T_*}^{T_c}
\frac{\Gamma(T)}{H(T)T^4}
e^{-\mathcal{I}(T)}
\,dT
\right]^{-1/3},
$
where $T_c$ is the critical temperature, $\Gamma(T)$ is the nucleation rate, and
$
\mathcal{I}(T)=
\frac{4\pi v_{\rm w}^{3}}{3}
\int_{T}^{T_c}
\frac{\Gamma(T')\,dT'}{T'^4H(T')}
\left(
\int_{T}^{T'}
\frac{d\bar{T}}{H(\bar{T})}
\right)^3$~\cite{Turner:1992tz,Kierkla:2022odc}.
} 
\begin{equation}
\label{eq-f1-f2}
f_1^\text{sw}
\simeq
0.068
\frac{H_{*,0}}{v_\text{w}}
\frac{\beta_{\text{GW}}}{H_*},
\qquad
f_2^\text{sw}
\simeq
0.17
\frac{H_{*,0}}{v_\text{w}\xi_\text{shell}}
\frac{\beta_{\text{GW}}}{H_*}
\max(v_{\text{w}},c_s),
\end{equation}
where $H_{*,0}$ is the Hubble parameter at the time of the phase transition,
redshifted to the present day, and $\xi_\text{shell}$ is the dimensionless sound
shell thickness~\cite{Espinosa:2010hh}.

Finally, the factor $\Upsilon$ in \autoref{eq:Omega-dbpl} incorporates the finite
lifetime of the sound-wave source, $\tau_{\text{sw}}$. This factor quantifies the 
reduction of the GW amplitude when the acoustic source is prematurely disrupted by 
turbulence or shocks instead of persisting for an entire Hubble time~\cite{Guo:2020grp}:
\begin{equation}
\label{eq-Upsilon}
  \Upsilon(H_* \tau_{\text{sw}})
  =
  1
  -
  \frac{1}{\sqrt{1+2H_* \tau_{\text{sw}}}}.
\end{equation}
In this context, $\tau_{\text{sw}}$ denotes the estimated decay timescale of sound 
waves into fully developed magnetohydrodynamic turbulence. It is bounded by the 
cosmological horizon and relates to the shock-formation timescale $\tau_\text{sh}$ via
\begin{equation}
H_* \tau_\text{sw}
=
\min\left[H_* \tau_\text{sh}, 1\right],
\end{equation}
where the explicit shock-formation time is 
\begin{equation}
\label{eq-Htaush}
H_* \tau_\text{sh} 
= \frac{H_* R_* }{\sqrt{\bar{U}^2_f} } 
\simeq
3.38 \, \frac{v_\text{w}}{\sqrt{K_\text{sw}}} \, \frac{H_*}{\beta_\text{GW}}.
\end{equation}
The parameter $\sqrt{\bar{U}^2_f}$ represents the root-mean-square fluid velocity, 
which satisfies the hydrodynamic relation $\bar{U}_f^2 = \frac{3}{4} K_{\text{sw}}$.

The detectability of the predicted GW signal at LISA is quantified
by the signal-to-noise ratio (SNR)~\cite{Caprini:2019egz}:
\begin{equation}
\label{eq:SNRLISA}
\text{SNR}
=
\sqrt{
\mathcal{T}
\int_{f_{\min}}^{f_{\max}} df
\left[
\frac{h^{2}\Omega_{\text{GW}}(f)}
{h^{2}\Omega_{\text{sens}}(f)}
\right]^2
},
\end{equation}
where $h^{2}\Omega_{\text{sens}}(f)$ is the nominal LISA sensitivity to a stochastic
background~\cite{Caprini:2019pxz,Babak:2021mhe}, and $f_{\min}$ and $f_{\max}$
define the sensitive frequency range. We take the mission duration to be
$\mathcal{T}=1.26\times10^8~\mathrm{s}$, corresponding to four years of observation
with an assumed duty cycle of unity~\cite{LISA:2017pwj}. Following standard
practice, we regard a GW signal as potentially detectable if
$\text{SNR}\gtrsim 10$~\cite{Caprini:2015zlo}.

\section{Scanning Methodology and Statistical Overview}
\label{sec-scan-method}

In this section, we present the global parameter-space scan and provide a
statistical overview of the viable SFOEWPT and GW configurations in the Normal
Scenario. To this end, we perform a comprehensive random scan over the 2HDM
parameter space within the following ranges:
\begin{equation}
\label{eq-scan-range}
\begin{alignedat}{3}
\mhh &\in [130,2000] \gev, & \quad
\ma &\in [15, 2000] \gev, & \quad
\mch &\in [80, 2000] \gev, \quad \\
|s_{\beta-\alpha}| &\in [0.9, 1], &\quad
\tb &\in [1, 50], &\quad
m_{12}^2 &\in [0, 2\times 10^6] \gev^2 .
\end{alignedat}
\end{equation}
We restrict the scan to positive values of $m_{12}^2$ to avoid the ``panic
vacuum'' scenario, in which a charged- or \textit{CP}-breaking minimum becomes
deeper than the desired electroweak-breaking vacuum~\cite{Barroso:2013awa,Ivanov:2015nea,Dorsch:2017nza}.

All sampled parameter points are first required to satisfy the standard
theoretical and experimental constraints defining a physically viable parameter point. On
the theoretical side, the scalar potential must be bounded from below~\cite{Ivanov:2006yq} and yield a stable electroweak vacuum~\cite{Ivanov:2008cxa,Barroso:2012mj,Barroso:2013awa}. We also enforce
tree-level unitarity in scalar--scalar scattering processes~\cite{Branco:2011iw,Arhrib:2000is} and require the Higgs quartic couplings to
remain within the perturbative regime~\cite{Chang:2015goa}. These requirements
are evaluated using the public code \package{2HDMC}~\cite{Eriksson:2009ws}.

Experimentally, the parameter points are required to satisfy electroweak
precision observables, flavor constraints, and collider limits.
We require agreement with the latest global fits to the electroweak oblique
parameters ($S,T,U$)~\cite{Peskin:1991sw,He:2001tp,Grimus:2008nb}, adopting
the updated Particle Data Group values $S=-0.04\pm0.10$, $T=0.01\pm0.12$,
and $U=-0.01\pm0.09$~\cite{ParticleDataGroup:2024cfk}, with their correlation
matrix included. 
To satisfy the stringent constraints from FCNCs, we apply
95\% confidence-level limits on rare $B$-meson decays, including
$B_{d,s}\to\mu^+\mu^-$~\cite{Haller:2018nnx}, $B\to X_s\gamma$~\cite{Haller:2018nnx,Arbey:2017gmh,Sanyal:2019xcp,Misiak:2017bgg},
$B\to K^*\gamma$~\cite{Belle:2017hum}, and
$B_s\to\phi\gamma$~\cite{Belle:2014sac}. Type-II and Type-Y are severely
constrained by these FCNC measurements~\cite{Haller:2018nnx,Misiak:2020vlo,Biekotter:2024ykp,Biekotter:2025fjx},
which impose a stringent lower bound on the charged Higgs mass,
$\mch\gtrsim 580 \gev$.

Finally, all viable points must evade exclusion limits from direct heavy-Higgs
searches at LEP, the Tevatron, and the LHC, while reproducing the precision
signal-strength measurements of the observed $125~\mathrm{GeV}$ SM-like Higgs
boson. Type-II and Type-X are also constrained by $\tau$-lepton-rich BSM
signals at the LHC~\cite{ATLAS:2017eiz,Kanemura:2011kx,Abe:2015oca,CMS:2018rmh,ATLAS:2020zms,Kim:2023lxc}.
This set of experimental constraints is imposed using \package{ScannerS}
version~2~\cite{Muhlleitner:2020wwk} interfaced with
\package{HiggsTools}-v1.2~\cite{Bahl:2022igd}.

Throughout this work, we classify the sampled points into three categories:
\begin{description}
\item[Physical parameter points:] parameter points that satisfy all theoretical
and experimental constraints described above. We obtain $10^6$ physical
parameter points for each 2HDM Yukawa type.
\item[SFOEWPT parameter points:] physical parameter points that pass the
cosmological viability criteria detailed in
Appendix~\ref{appendix-BSMPT-selection} and satisfy $\xi_p\geq 1$. Their
thermal histories are evaluated using \package{BSMPT} version~3.1.8.
\item[GW parameter points:] SFOEWPT parameter points that additionally yield a
potentially observable stochastic GW signal, requiring
$\mathrm{SNR}\geq 10$ at LISA for a four-year mission. The corresponding GW
signals are computed using \package{BSMPT} version~3.1.8.
\end{description}
This classification allows us to compare the thermal history, phase-transition
dynamics, and GW phenomenology across the four Yukawa types and the two resummation schemes.

\begin{table}[t]
\centering
\caption{
Comparative statistics of the SFOEWPT and GW parameter configurations under the
Parwani and AE resummation schemes. The metrics are tabulated across the
single-step ($\trans{1}{1}$) channel and the second stage of the two-step
($\trans{2}{2}$) SFOEWPT sequence for all four Yukawa types of the 2HDM, listing
the total number of SFOEWPT points ($\nsfo$), the maximum electroweak order
parameter at the percolation temperature ($\xi_p^{\rm max}$), the number of GW
parameter points ($N_\text{GW}$), and the maximum signal-to-noise ratio
($\rm{SNR}^{\rm max}$).\\[2pt]
}
\label{tab-comparison-P-AE}
\setlength{\tabcolsep}{10pt} 
\renewcommand{\arraystretch}{1.3} 
\begin{tabular}{|l|l|l|r|r|r|r|}
\hline
Transition & Scheme & Quantity & Type-I & Type-II & Type-X & Type-Y \\ \hline\hline
\multirow{8}{*}{$\trans{1}{1}$} & \multirow{4}{*}{Parwani} & $\nsfo$ & 66,371 & 105,677 & 95,737 & 91,207 \\ \cline{3-7} 
 &  & $\xi_p^{\rm max}$ & 3.94 & 3.87 & 3.90 & 3.87 \\ \cline{3-7} 
 &  & $N_\text{GW}$ & 1,789 & 63 & 670 & 61 \\ \cline{3-7} 
 &  & $\rm{SNR}^{\rm max}$ & 95.14 & 19.38 & 60.41 & 19.38 \\ \cline{2-7} 
 & \multirow{4}{*}{AE} & $\nsfo$ & 12,824 & 230 & 1,658 & 278 \\ \cline{3-7} 
 &  & $\xi_p^{\rm max}$ & 5.62 & 1.61 & 4.13 & 1.75 \\ \cline{3-7} 
 &  & $N_\text{GW}$ & 110 & 0 & 1 & 0 \\ \cline{3-7} 
 &  & $\rm{SNR}^{\rm max}$ & 195.11 & -- & 54.27 & -- \\ \hline\hline
\multirow{4}{*}{$\trans{2}{2}$} & \multirow{2}{*}{Parwani} & $\nsfo$ & 14 & 0 & 3 & 0 \\ \cline{3-7} 
 &  & $\xi_p^{\rm max}$ & 1.58 & -- & 1.29 & -- \\ \cline{2-7} 
 & \multirow{2}{*}{AE} & $\nsfo$ & 5 & 0 & 1 & 0 \\ \cline{3-7} 
 &  & $\xi_p^{\rm max}$ & 1.47 & -- & 1.10 & -- \\ \hline
\end{tabular}
\end{table}

To establish a baseline, we begin with a global statistical overview of the 
viable SFOEWPT and GW parameter configurations before analyzing specific parameter 
space distributions and detailed hydrodynamic properties in subsequent sections. 
Table~\ref{tab-comparison-P-AE} summarizes the total counts of successful SFOEWPT 
and GW configurations, along with the maximum achieved values for $\xi_p$ and the 
four-year LISA SNR, across all four 2HDM types. For a direct comparison of the 
theoretical uncertainties, the results are presented separately for the Parwani 
and AE resummation schemes.

Several phenomenological trends emerge from
Table~\ref{tab-comparison-P-AE}. First, one-step transitions
($\trans{1}{1}$) overwhelmingly dominate the SFOEWPT parameter space across all
four Yukawa types in the Normal Scenario. Configurations in which the second stage of a two-step transition
($\trans{2}{2}$) is strongly first-order are exceedingly rare. Under the
Parwani scheme, only 14 such points are found in Type-I and only 3 in Type-X,
while none appear in Type-II or Type-Y. The AE scheme suppresses these transitions
even further, reducing their number to roughly one third of the Parwani yield.
All other multi-step SFOEWPT topologies, such as $\trans{1}{2}$ or any
transition within a three-step sequence ($\trans{i}{3}$), are absent from our
Normal Scenario dataset irrespective of the resummation scheme.

This behavior contrasts sharply with the Inverted Scenario. As demonstrated in
our previous study of the Type-I 2HDM using the AE resummation scheme~\cite{Lee:2025hgb}, the Inverted Scenario accommodates a much richer diversity
of thermal histories.
Out of $2.36\times10^6$ physical parameter
points, the yields are
\begin{equation}
\label{eq-IS-N_SFO}
\text{Inverted Scenario in Type-I: }
\begin{cases}
    \nsfo = 5343 & \text{for }  \trans{1}{1}, \\
    \nsfo = 4486 & \text{for }  \trans{2}{2}.
\end{cases}
\end{equation}
Exotic transition histories such as $\trans{1}{2}$ and $\trans{3}{3}$ are rare
but nonzero in the Inverted Scenario. Thus, \autoref{tab-comparison-P-AE} shows that the
Normal mass hierarchy strongly suppresses multi-step electroweak phase
transitions.

The second prominent feature of Table~\ref{tab-comparison-P-AE} is the
substantial discrepancy between the Parwani and AE resummation schemes, visible
in both the SFOEWPT and GW parameter spaces. This discrepancy represents a
significant source of theoretical uncertainty. For the SFOEWPT parameter points,
the Parwani scheme yields a large number of successful configurations,
$\mathcal{O}(10^4)$--$\mathcal{O}(10^5)$ per type. This broad parameter space
supports efficient SFOEWPTs across all four Yukawa types, with maximum order
parameters reaching $\xi_p^{\rm max}\sim 3.9$. For single-step transitions,
Type-II yields the largest number of SFOEWPT points, followed by Type-X,
Type-Y, and Type-I.

By contrast, the AE scheme drastically suppresses the total 
number of viable SFOEWPT parameter points. For single-step transitions, the AE 
yield is reduced to approximately $19.3\%$ of the Parwani yield in Type-I, and 
dwindles to a mere $\mathcal{O}(0.1)\%$--$\mathcal{O}(1)\%$ across the other three 
Yukawa structures. Furthermore, the AE prescription introduces a pronounced model 
dependence regarding the maximum strength of the transition. While Type-I and 
Type-X retain large maximum values for the order parameter, yielding 
$\xi_p^{\rm max}=5.62$ and $\xi_p^{\rm max}=4.13$ respectively, Type-II and 
Type-Y exhibit significantly constrained transition strengths, bounded at 
$\xi_p^{\rm max}=1.61$ and $\xi_p^{\rm max}=1.75$ respectively.
 
An even more profound dependence on both the resummation 
schemes and the Yukawa types emerges when examining the viable GW parameter configurations. Under the 
Parwani prescription, Type-I accommodates a substantial GW sample with $N_\text{GW}=1789$. 
While significantly more suppressed by comparison, the remaining Yukawa structures 
still yield statistically non-negligible samples ranging from $N_\text{GW}=\mathcal{O}(10)$ 
to $\mathcal{O}(10^2)$ points. Under the AE scheme, however, the overall yield drops 
sharply to $110$ in Type-I and essentially vanishes across the other three Yukawa 
types. Crucially, despite this drastic reduction in the absolute count of GW parameter 
points relative to the total SFOEWPT sample size, the predicted GW 
signals themselves can remain exceptionally strong. Type-I achieves a maximum 
SNR of approximately 95 under the Parwani scheme, which 
further escalates to 195 under the AE framework. Consequently, while the viable 
GW parameter space is severely constrained compared to the broad SFOEWPT boundaries, 
upcoming LISA observations retain an excellent sensitivity to probe the strongest 
phase transitions predicted within the 2HDM landscape.

\section{Comparison of Thermal Resummation Schemes in the SFOEWPT Parameter Space}
\label{sec:PvsAE}

The pronounced discrepancy in the total counts of viable SFOEWPT and GW parameter 
points induced by the choice of resummation scheme naturally raises the 
question of how their distributions differ within the underlying 2HDM landscape. 
If the preferred parameter regions vary substantially with the thermal resummation 
prescription, the corresponding search strategies at future collider experiments 
will inevitably inherit significant theoretical uncertainties. To address this 
issue, we systematically compare the distributions of the surviving SFOEWPT points between
the Parwani and AE schemes, focusing on the Type-I and Type-II models.

\begin{figure}[t]
\centering
\includegraphics[width=\textwidth]{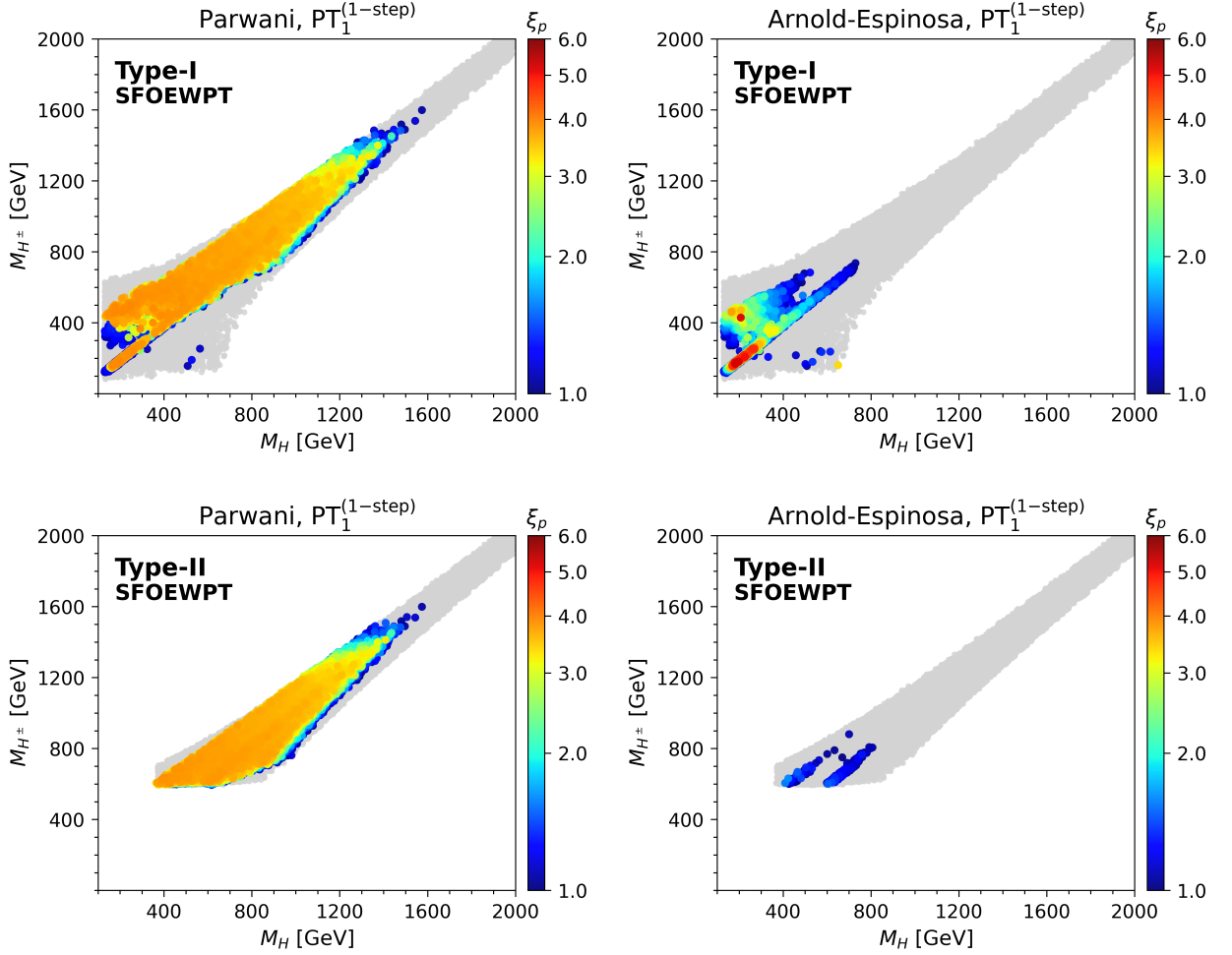}
\caption{%
Distribution of SFOEWPT parameter points in the $(M_H, M_{H^\pm})$ plane for the single-step transition scenario in Type-I (upper panels) and Type-II (lower panels). The left and right panels compare results obtained using the Parwani and AE resummation schemes, respectively. The color scale denotes the order parameter $\xi_p$. The underlying gray points represent the initial physical parameter points satisfying all theoretical and experimental constraints. To highlight regions of stronger phase transitions, points with higher $\xi_p$ are plotted on top of those with lower values.
}
\label{fig-PvsAE-McH-MH-xip}
\end{figure}

\autoref{fig-PvsAE-McH-MH-xip} displays the SFOEWPT parameter points in the
$(M_H,M_{H^\pm})$ plane for Type-I (upper panels) and Type-II (lower panels).
Since multi-step SFOEWPTs are exceedingly rare in the Normal Scenario (see
\autoref{tab-comparison-P-AE}), we focus exclusively on the single-step
transition. The left and right panels compare the Parwani and AE resummation
schemes, respectively, with the color scale representing the order parameter
$\xi_p$. The underlying gray points denote the physical parameter space that
satisfies all theoretical and experimental constraints before imposing the
cosmological viability conditions and the additional requirement $\xi_p>1$.

The most striking feature of \autoref{fig-PvsAE-McH-MH-xip} is the pronounced
difference between the parameter-space distributions obtained with the Parwani
and AE resummation schemes. Although both prescriptions impose strict upper
bounds on $M_H$ and $M_{H^\pm}$ that are absent in the full set of physical
parameter points, the scale of the resulting mass cutoff differs
substantially. Under the Parwani prescription, the viable SFOEWPT parameter
space extends up to approximately $1.6\tev$ in both Type-I and Type-II. In
contrast, the AE scheme yields significantly stronger restrictions, limiting
the spectrum to $M_{H,H^\pm}\lesssim 750\gev$ in Type-I and
$M_{H,H^\pm}\lesssim 850\gev$ in Type-II.

Remarkably, these upper bounds on the heavy-scalar masses provide encouraging
prospects for collider searches of cosmologically viable 2HDMs. Even the more
conservative Parwani bound of $\sim 1.6\tev$ lies well within the projected
reach of future multi-TeV muon colliders~\cite{Palmer:1996gs,Ankenbrandt:1999cta,Capdevilla:2020qel,Bandyopadhyay:2021pld,Sen:2021fha,Asadi:2021gah,Huang:2021nkl}. By contrast, the corresponding lower bounds on
the heavy Higgs masses are largely insensitive to the choice of resummation
scheme and are instead primarily determined by the Yukawa structure.

Another notable feature is the highly fragmented topology produced by the AE
scheme, particularly in Type-II. In this case, the SFOEWPT parameter points do
not populate continuous regions of parameter space, but instead appear as
isolated clusters separated by irregular voids. Such fragmentation may arise
either from numerical instabilities in the evaluation of the finite-temperature
effective potential or from an extreme sensitivity of the transition dynamics
to the underlying model parameters, which becomes exposed once the strong
transition requirement, $\xi_p>1$, is imposed.

\begin{figure}[t]
\centering
\includegraphics[width=0.6\textwidth]{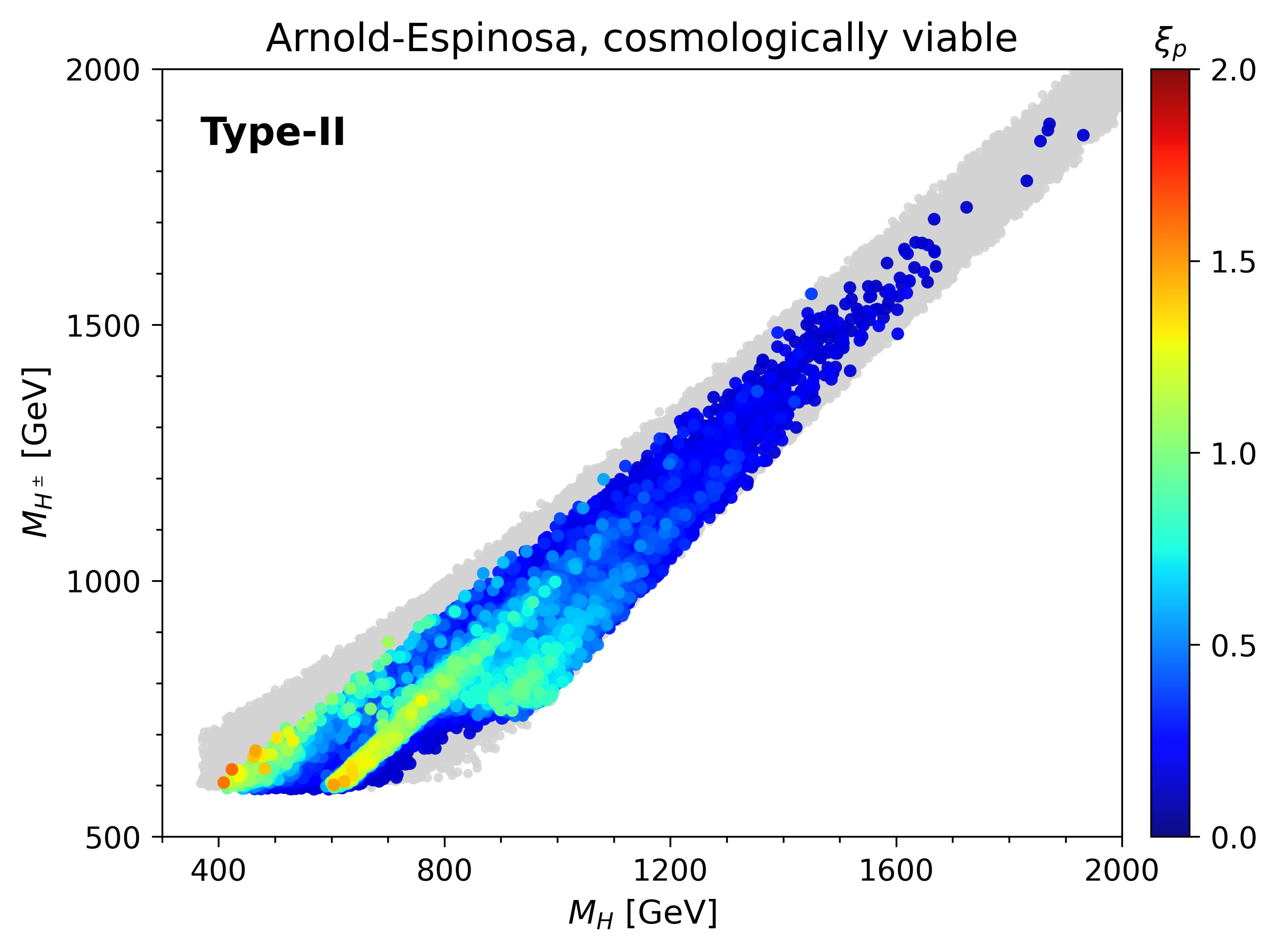}
\caption{%
Distribution of cosmologically viable parameter points in the $(M_H, M_{H^\pm})$ plane for the single-step transition scenario in Type-II under the AE resummation scheme, without applying the strong phase transition criterion $\xi_p>1$. The color scale denotes the order parameter $\xi_p$, and the underlying gray points represent the physical parameter points.
}
\label{fig-type2-AE-McH-MH-no-strong}
\end{figure}

To identify the origin of this fragmentation,
\autoref{fig-type2-AE-McH-MH-no-strong} shows the corresponding distribution of
cosmologically viable Type-II parameter points under the AE scheme, without
imposing the strong-transition requirement $\xi_p>1$. Out of $10^6$ physical
parameter points, $3.80\times 10^4$ configurations satisfy the cosmological
viability conditions in the AE scheme. Once the $\xi_p>1$ requirement is
removed, the parameter points recover a smooth and continuous distribution, with
no visible fragmentation or irregular voids. This demonstrates that the AE
prescription itself does not suffer from numerical instabilities in the
evaluation of the finite-temperature effective potential.

Instead, the comparison between the distributions obtained with and without the
$\xi_p>1$ requirement points to a limitation of the AE framework. The effective
potential evaluated with the AE prescription exhibits extreme parametric
sensitivity: small variations in the heavy-Higgs masses can induce large
fluctuations in the order parameter $\xi_p$. Applying the sharp threshold
$\xi_p>1$ to such a sensitive landscape then truncates the parameter space in a
highly nonuniform manner, generating the fragmented structures observed in the
SFOEWPT distributions.

\begin{figure}[t]
\centering
\includegraphics[width=\textwidth]{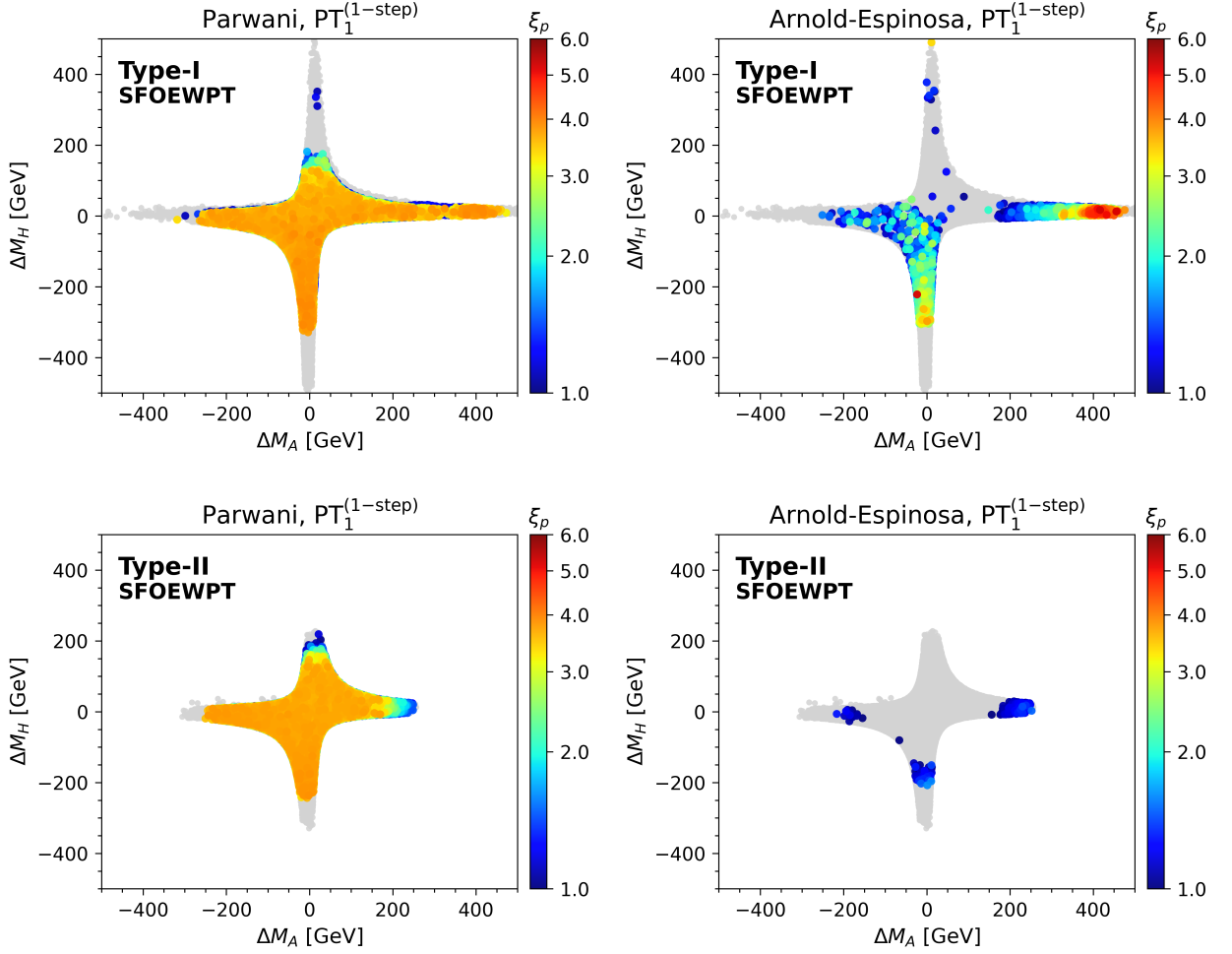}
\caption{%
Distribution of SFOEWPT parameter points in the $(\Delta M_A, \Delta M_{H})$ plane for the single-step transition scenario in Type-I (upper panels) and Type-II (lower panels), where $\Delta M_i = M_i - M_{H^\pm}$. The left and right panels compare results obtained using the Parwani and AE resummation schemes, respectively.
The color scale denotes the order parameter $\xi_p$. The underlying gray points represent the initial physical parameter points.
}
\label{fig-PvsAE-DMA-DMH-xip}
\end{figure}

Another pronounced difference between the two resummation schemes appears in the
distributions of the heavy-scalar mass splittings.
\autoref{fig-PvsAE-DMA-DMH-xip} shows the SFOEWPT parameter points in the
$(\Delta M_A,\Delta M_H)$ plane, where
$\Delta M_i\equiv M_i-M_{H^\pm}$. We again restrict the discussion to
single-step transitions and compare Type-I (upper panels) and Type-II (lower
panels) under the Parwani (left panels) and AE (right panels) prescriptions.

Under the Parwani scheme, the SFOEWPT parameter points 
form broad and continuous
distributions centered around the custodial mass-degenerate region,
$M_H\simeq M_A\simeq M_{H^\pm}$, with both positive and negative mass splittings
allowed. Moreover, larger values of $\xi_p$ preferentially populate the
vicinity of the custodial region, indicating that stronger phase transitions favor
approximate mass degeneracy.

By contrast, the AE scheme produces markedly asymmetric distributions. In
Type-I, this asymmetry is manifested by the strong suppression of parameter
points with $\Delta M_H>0$. In addition, the largest values of $\xi_p$ are
confined to a narrow region around $\Delta M_H\simeq 0$ and
$\Delta M_A\simeq 430\gev$. The asymmetry becomes even more severe in Type-II,
where the custodial mass-degenerate region is almost entirely excluded.
Together with the fragmented point distributions discussed above, these
features provide further evidence that the AE prescription induces an extreme
sensitivity of the finite-temperature effective potential to the heavy-scalar
spectrum. Consequently, if future collider measurements were to indicate an
approximately custodial-degenerate heavy-scalar spectrum, the AE description of
cosmologically viable SFOEWPTs would be strongly disfavored.

\begin{figure}[t]
\centering
\includegraphics[width=\textwidth]{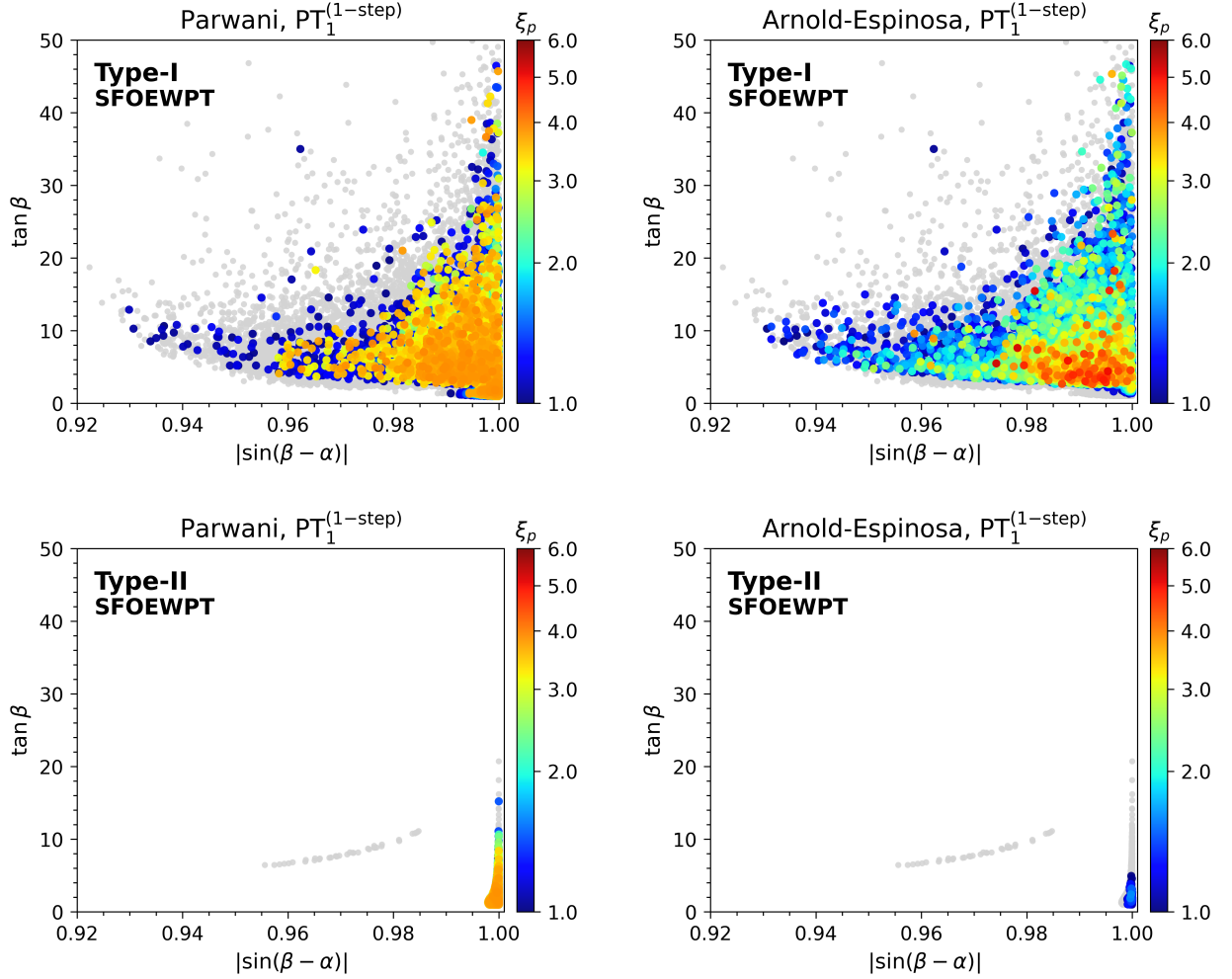}
\caption{%
Distribution of SFOEWPT parameter points in the $(|s_{\beta-\alpha}|, \tb)$ plane for the single-step transition scenario in Type-I (upper panels) and Type-II (lower panels).
The left and right panels compare results obtained using the Parwani and AE resummation schemes, respectively.
The color scale denotes the order parameter $\xi_p$. 
}
\label{fig-PvsAE-tb-abs_sba-xip}
\end{figure}

In contrast to the strong scheme dependence observed in the heavy-scalar mass
spectra, the scalar mixing parameters are remarkably insensitive to the choice
of thermal resummation. \autoref{fig-PvsAE-tb-abs_sba-xip} shows the SFOEWPT
parameter points in the $(|s_{\beta-\alpha}|,\tb)$ plane. The overall distributions obtained with the Parwani and AE schemes are highly
similar. Even in Type-II, where the AE prescription yields a slightly narrower
range of $\tb$, the global structure of the parameter space remains essentially
unchanged.

In addition, the extreme sensitivity of the AE prescription to the $\xi_p$
requirement, which manifests itself through fragmented structures in the
heavy-scalar mass planes, is absent in the scalar mixing parameter space. Even
after imposing the stringent condition $\xi_p>1$, the AE parameter points
remain smoothly distributed, with no evidence of irregular voids or isolated
clusters.

\begin{figure}[t]
\centering
\includegraphics[width=\textwidth]{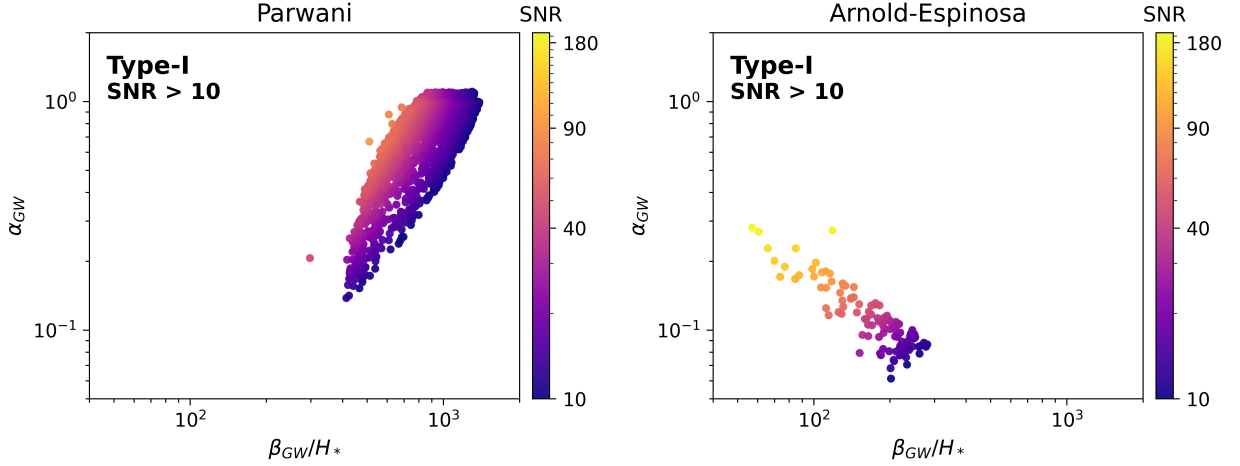}
\caption{%
Distribution of the viable GW parameter points in the $(\beta_{\text{GW}}/H_*,\alpha_{\text{GW}})$ plane for the Type-I model under the Parwani (left panel) and AE (right panel) resummation schemes. The color scale denotes the four-year LISA signal-to-noise ratio.
}
\label{fig-PvsAE-alpha-beta-SNR}
\end{figure}

Finally, we examine the impact of the thermal-resummation prescription on the
macroscopic phase-transition parameters that govern the GW spectrum. In
\autoref{fig-PvsAE-alpha-beta-SNR}, we project the viable GW parameter points
onto the $(\beta_{\rm GW}/H_*,\alpha_{\rm GW})$ plane, comparing the
distributions obtained with the Parwani (left panel) and
AE (right panel) schemes for the Type-I model. The color scale
indicates the corresponding four-year LISA SNR. This
comparison reveals that the choice of thermal-resummation prescription does not
merely shift the viable parameter space, but qualitatively reshapes the
macroscopic phase-transition dynamics.

There are several phenomenal features in \autoref{fig-PvsAE-alpha-beta-SNR}.
First, the transition strength parameter $\alpha_{\rm GW}$ is substantially
enhanced in the Parwani scheme, reaching values of
$\mathcal{O}(1)$. In contrast, the AE scheme restricts
$\alpha_{\rm GW}$ to values below approximately $0.3$. 
The enhancement of $\alpha_{\rm GW}$ in the Parwani scheme indicates a larger
vacuum-energy release during the phase transition. Since the dominant GW source
is the acoustic motion of the plasma, this directly enhances the sound-wave
contribution and yields larger GW amplitudes.
 This provides a natural explanation for the
significantly larger number of detectable GW parameter points obtained in the
Parwani scheme compared with the AE scheme, as summarized in
\autoref{tab-comparison-P-AE}.

Second, the two prescriptions populate markedly different ranges of the inverse-duration parameter. The Parwani scheme yields
$\beta_{\rm GW}/H_*\gtrsim 300$, extending up to
$\mathcal{O}(10^3)$, whereas the AE scheme produces values
entirely below approximately $300$, reaching as low as
$\sim 60$. Since $\beta_{\rm GW}/H_*$ is inversely related to the duration of
the phase transition, this indicates that the Parwani scheme favors
substantially shorter transition durations than the AE scheme.

Third, the two prescriptions yield qualitatively different structures in the
$(\beta_{\rm GW}/H_*,\alpha_{\rm GW})$ plane. The Parwani scheme exhibits a
mild positive trend, whereas the AE scheme tends toward an anti-correlation
between the two quantities. This difference is particularly noteworthy because
it arises within the same underlying BSM model. The results of
\autoref{fig-PvsAE-alpha-beta-SNR} therefore demonstrate that the choice of
thermal-resummation prescription does not merely shift the viable parameter
space, but can qualitatively reshape the macroscopic phase-transition
parameters relevant for GW production. This sensitivity originates from the
fact that $\alpha_{\rm GW}$ and $\beta_{\rm GW}/H_*$ are determined by the
finite-temperature effective potential and the associated bounce action, both
of which are directly affected by the thermal resummation of daisy diagrams.
Consequently, the observed differences reveal a substantial theoretical
uncertainty associated with the treatment of thermal resummation in GW analyses.

In summary, the choice of thermal-resummation prescription has a substantial
impact on both the viable SFOEWPT parameter space and the resulting
macroscopic phase-transition parameters relevant for GW production. In
particular, the AE prescription exhibits an extreme sensitivity to the
$\xi_p>1$ selection criterion, leading to fragmented parameter-space
distributions. By contrast, the Parwani scheme yields numerically stable and
structurally continuous SFOEWPT parameter spaces across all parameter
projections. Furthermore, recent studies have shown that the Parwani
prescription closely reproduces the results of the gauge-independent
partial-dressing framework for multi-field scalar
potentials~\cite{Bahl:2024ykv,Bittar:2025lcr}. Therefore, we adopt the
Parwani prescription throughout the remainder of this work.

\section{Characteristics of SFOEWPT Parameter Points under the Parwani Scheme}
\label{sec:characteristics:SFOEWPT:Parwani}

We now turn to one of the main results of this work: a comparative study of
cosmologically viable SFOEWPT parameter points across the four Yukawa
realizations of the 2HDM, namely Types I, II, X, and Y. In this section, we
investigate the characteristic features of the SFOEWPT parameter space under
the Parwani prescription.

\begin{figure}[t]
\centering
\includegraphics[width=\textwidth]{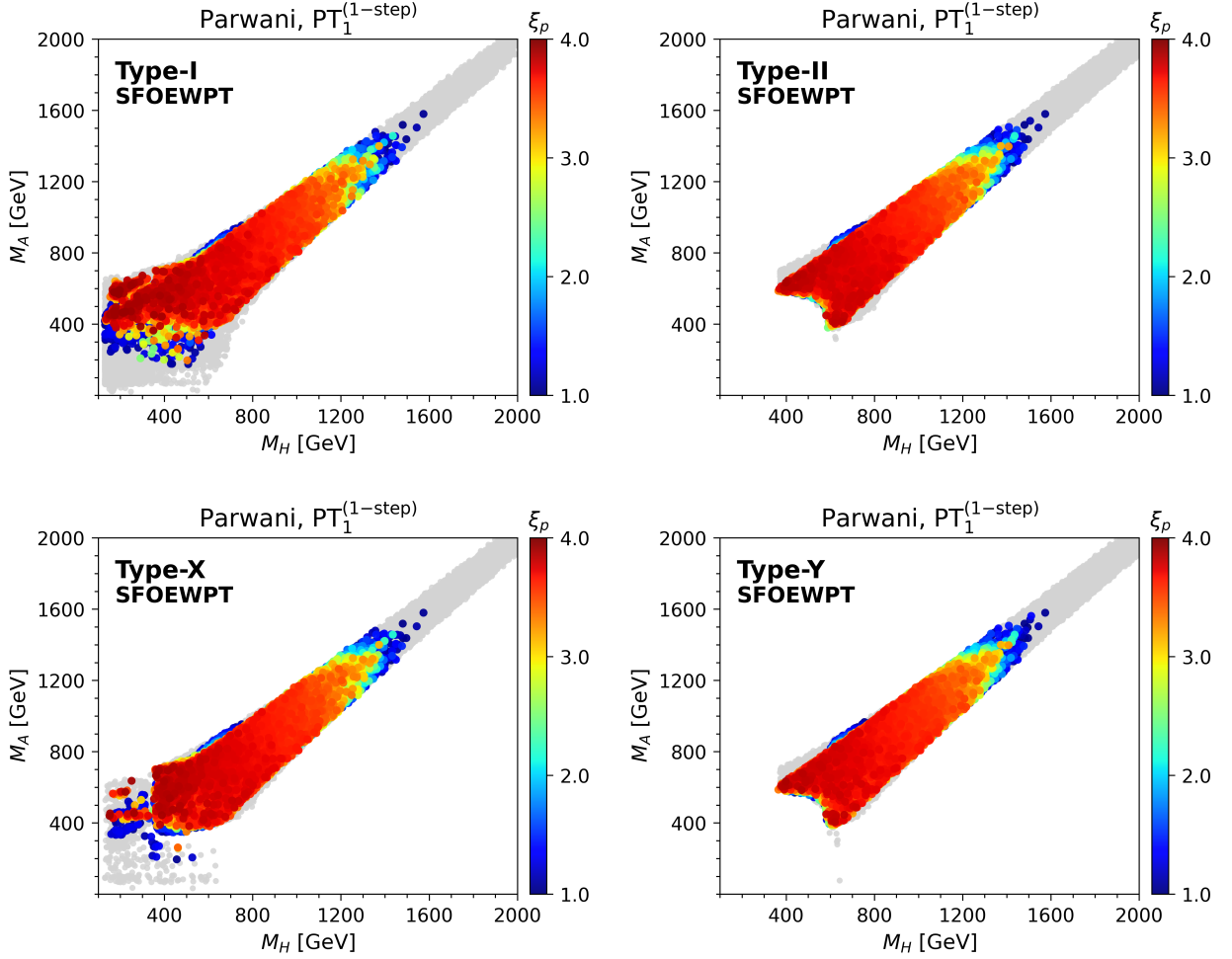}
\caption{%
Distribution of SFOEWPT parameter points in the $(M_H, M_A)$ plane for the one-step transition scenario across the four 2HDM types under the Parwani resummation scheme. The color scale denotes the phase transition order parameter $\xi_p$. The sub-panels correspond to Type-I (upper left), Type-II (upper right), Type-X (lower left), and Type-Y (lower right).
}
\label{fig-4types-MA-MH-xip}
\end{figure}

We begin by examining the heavy-Higgs mass spectrum within the SFOEWPT
parameter space. \autoref{fig-4types-MA-MH-xip} shows the distribution of
dominant single-step transition points in the $(M_H,M_A)$ plane for the four
Yukawa types. The color scale denotes $\xi_p$.

The most striking feature is the emergence of a common upper bound on the heavy
neutral Higgs masses, $M_{H,A}\lesssim 1.6\tev$, across all four Yukawa types.
This behavior is consistent with the well-established field-theoretic
expectation that the decoupling limit disfavors an SFOEWPT: as the heavy-Higgs
masses increase, the order parameter $\xi_p$ decreases
asymptotically~\cite{Quiros:1999jp,Su:2020pjw,Niemi:2021qvp,Bahl:2024ykv,Masina:2025pnp}.
Nevertheless, the upper bound obtained with the Parwani prescription is
considerably higher than the corresponding value of approximately $1\tev$
reported using the AE scheme~\cite{Basler:2016obg,Su:2020pjw}, highlighting the
significant impact of thermal resummation on the heavy-Higgs mass range allowed
by the SFOEWPT condition.

In contrast to the nearly universal upper bounds, the lower boundaries of the
SFOEWPT parameter space exhibit a pronounced dependence on the Yukawa type. For
Type-II and Type-Y, the lower limits are dictated almost entirely by existing
experimental constraints rather than by the thermodynamics of the phase
transition, as indicated by the close agreement between the lower boundaries of
the physical parameter space, shown by the gray points, and those of the
SFOEWPT parameter space. In particular, indirect flavor measurements, most
notably $B\to X_s\gamma$, impose the stringent constraint
$\mch\gtrsim 580\gev$~\cite{Haller:2018nnx,Misiak:2020vlo,Biekotter:2024ykp,Biekotter:2025fjx}.
Electroweak precision observables further require $M_H$ and $M_A$ to remain
close to $\mch$ in order to preserve approximate custodial symmetry.
Consequently, imposing the SFOEWPT condition introduces little additional
restriction at the low bounds on $M_H$ and $M_A$ in Type-II and Type-Y.

In Type-I and Type-X, however,  the lower bound on $M_A$ is determined
primarily by the thermodynamics of the electroweak phase transition. Since
flavor constraints are much weaker in these two Yukawa types, requiring a
successful SFOEWPT removes a substantial portion of the low-$M_A$ region,
resulting in the cosmological bounds $M_A\gtrsim 197\gev$ in Type-I and
$M_A\gtrsim 206\gev$ in Type-X. The corresponding distribution of $M_H$,
however, remains comparatively insensitive to the SFOEWPT condition.

\begin{figure}[t]
\centering
\includegraphics[width=\textwidth]{fig-4types-tb-MHc-xip}
\caption{%
Distribution of SFOEWPT parameter points in the $(M_{H^\pm}, \tb)$ plane for the single-step transition scenario across the four 2HDM types under the Parwani resummation scheme. The color scale denotes the order parameter $\xi_p$. The subpanels correspond to Type-I (upper left), Type-II (upper right), Type-X (lower left), and Type-Y (lower right).
}
\label{fig-4types-tb-MHc-xip}
\end{figure}

We next examine the SFOEWPT parameter space in the
$(M_{H^\pm},\tb)$ plane, shown in
\autoref{fig-4types-tb-MHc-xip}. This projection is of particular interest
because the combination of $\tb$ and $M_{H^\pm}$ plays a central role in many
of the flavor constraints on the
2HDM~\cite{Haller:2018nnx,Misiak:2020vlo,Biekotter:2024ykp,Biekotter:2025fjx}.

Consistent with the behavior observed for the neutral Higgs masses, the
SFOEWPT condition truncates the decoupling regime, imposing a nearly universal
upper bound, $M_{H^\pm}\lesssim 1.6~\tev$, across all four Yukawa types.
 By contrast, the
lower bound on $M_{H^\pm}$ remains largely unchanged, as it is already
determined by flavor constraints. Likewise, the allowed range of $\tb$ is only
mildly affected by the SFOEWPT requirement and continues to span most of the
physically viable parameter space.

An interesting feature of this projection is that the geometric extent of the
SFOEWPT parameter space does not directly track the total number of successful
SFOEWPT points reported in \autoref{tab-comparison-P-AE}. Although Type-I
contains the fewest SFOEWPT points, it occupies the largest region in the
$(M_{H^\pm},\tb)$ plane, followed closely by Type-X. Conversely, despite
yielding the largest number of SFOEWPT points, Type-II is confined to a much
smaller region of parameter space. This observation highlights that the number
of viable SFOEWPT points alone does not fully characterize the structure of the
allowed parameter space. From a phenomenological perspective, the geometric
extent of the viable region provides complementary information, indicating the
range of charged-Higgs masses and $\tb$ values over which future searches may
probe cosmologically viable realizations of the 2HDM.

\begin{figure}[t]
\centering
\includegraphics[width=\textwidth]{fig-4types-tb-sba-xip}
\caption{%
Distribution of SFOEWPT parameter points in the $(s_{\beta-\alpha}, \tb)$ plane for the one-step transition scenario across the four 2HDM types under the Parwani resummation scheme. The color scale denotes the phase transition order parameter $\xi_p$. The subpanels correspond to Type-I (upper left), Type-II (upper right), Type-X (lower left), and Type-Y (lower right).
}
\label{fig-4types-tb-sba-xip}
\end{figure}

We next examine the scalar mixing structure of the SFOEWPT parameter space by
projecting the successful configurations onto the
$(s_{\beta-\alpha},\tb)$ plane, as shown in
\autoref{fig-4types-tb-sba-xip}. The mixing parameter
$s_{\beta-\alpha}$ governs the Higgs alignment limit. From the decomposition of
$h_{\rm SM}$ in terms of $h$ and $H$ (see \autoref{eq-hsm-def}), the two
branches $s_{\beta-\alpha}=\pm1$ both correspond to exact Higgs alignment.
Along the $s_{\beta-\alpha}=-1$ branch, the top-quark Yukawa modifier,
$\kappa_t=c_\alpha/s_\beta=s_{\beta-\alpha}+c_{\beta-\alpha}/\tb$,
also approaches $-1$, thereby preserving the relative sign between the gauge and
top-quark Yukawa couplings and maintaining the SM-like interference pattern in
the diphoton decay.

The distributions in the $(s_{\beta-\alpha},\tb)$ plane exhibit a pronounced
dependence on the Yukawa type. Only Type-I permits sizable departures from the
alignment limit, with the viable interval
$s_{\beta-\alpha}\in[-1,-0.930]\cup[0.975,1]$. Within this region, larger
values of $\xi_p$ show a mild preference for exact alignment. Furthermore, the
Type-I distribution is noticeably asymmetric: for $\tb\gtrsim20$, viable
parameter points almost exclusively populate the positive-$s_{\beta-\alpha}$
branch. By contrast, Type-II and Type-Y strongly favor the alignment limit,
restricting deviations from $|s_{\beta-\alpha}|=1$ to the level of
$\mathcal{O}(10^{-3})$.

Another notable feature of
\autoref{fig-4types-tb-sba-xip} is the appearance of extended arm-like
structures in the physical parameter space (gray points) of Type-II, Type-X,
and Type-Y. These branches originate from the wrong-sign Yukawa regime,
defined by the condition
$\mathrm{sign}(\kappa_V)\kappa_{f_{\rm w}}<0$~\cite{Ferreira:2014naa,Coyle:2018ydo,Ferreira:2017bnx,Su:2019ibd,Han:2020zqg},
where $f_{\rm w}$ denotes the fermion species capable of realizing a wrong-sign
Yukawa coupling: the $b$ quark and $\tau$ lepton in Type-II, the $\tau$ lepton
in Type-X, and the $b$ quark in Type-Y. Using the coupling relations in
\autoref{tab-Yukawa}, the corresponding Yukawa modifier is
\[
\kappa_{f_{\rm w}}
=- \frac{s_\alpha}{c_\beta}
=s_{\beta-\alpha}-c_{\beta-\alpha}\tb .
\]
For positive $s_{\beta-\alpha}$, the wrong-sign limit
$\kappa_{f_{\rm w}}\simeq-1$ is realized approximately along the trajectory
\[
\tb\simeq\frac{1+s_{\beta-\alpha}}{c_{\beta-\alpha}},
\]
which accurately reproduces the geometry of these arm-like structures.
Remarkably, after imposing the SFOEWPT condition, only Type-X retains a
substantial portion of the wrong-sign branch, allowing
$s_{\beta-\alpha}$ to deviate from exact alignment down to approximately
0.97.

\begin{figure}[t]
\centering
\includegraphics[width=\textwidth]{fig-4types-m12sq-tb-xip}
\caption{%
Distribution of SFOEWPT parameter points in the $(\tb, m_{12}^2)$ plane for the single-step transition scenario across the four 2HDM types under the Parwani resummation scheme. The color scale denotes the order parameter $\xi_p$. The subpanels correspond to Type-I (upper left), Type-II (upper right), Type-X (lower left), and Type-Y (lower right).
}
\label{fig-4types-m12sq-tb-xip}
\end{figure}

Having examined the Higgs mass spectrum and mixing parameters, we next turn to
the soft-breaking parameter $m_{12}^2$, which controls the strength of the soft
breaking of the $Z_2$ symmetry. \autoref{fig-4types-m12sq-tb-xip} displays the
SFOEWPT parameter points in the $(\tb,m_{12}^2)$ plane for the single-step
transition scenario across the four Yukawa types, with the color code of $\xi_p$.

The first notable feature is that the SFOEWPT requirement reduces the maximum
allowed value of $m_{12}^2$ by approximately a factor of two.
 The
resulting upper bound,
$m_{12,\max}^2 \simeq 10^6\gev^2$,
is nearly identical across all four Yukawa types and is consistently realized
at the smallest values of $\tb$. By contrast, the lower bound on $m_{12}^2$
exhibits a moderate type dependence, remaining at
$\mathcal{O}(10^3)\gev^2$ in Type-I and Type-X, while increasing to
$\mathcal{O}(10^4)\gev^2$ in Type-II and Type-Y. These results indicate that a successful SFOEWPT favors substantial, nonzero
soft breaking of the discrete $Z_2$ symmetry.

We further observe that larger values of $\xi_p$ systematically favor smaller
values of both $m_{12}^2$ and $\tb$. The preference for low $\tb$ implies that
the viable SFOEWPT parameter space disfavors a substantial suppression of the
top-quark Yukawa coupling. This behavior reflects the important role played by
top-quark-induced thermal corrections in shaping the finite-temperature
effective potential. As $\tb$ increases, the top Yukawa coupling becomes
progressively suppressed, reducing its impact on the thermal evolution of the
potential and making it more difficult to realize an SFOEWPT.

\begin{figure}[t]
\centering
\includegraphics[width=\textwidth]{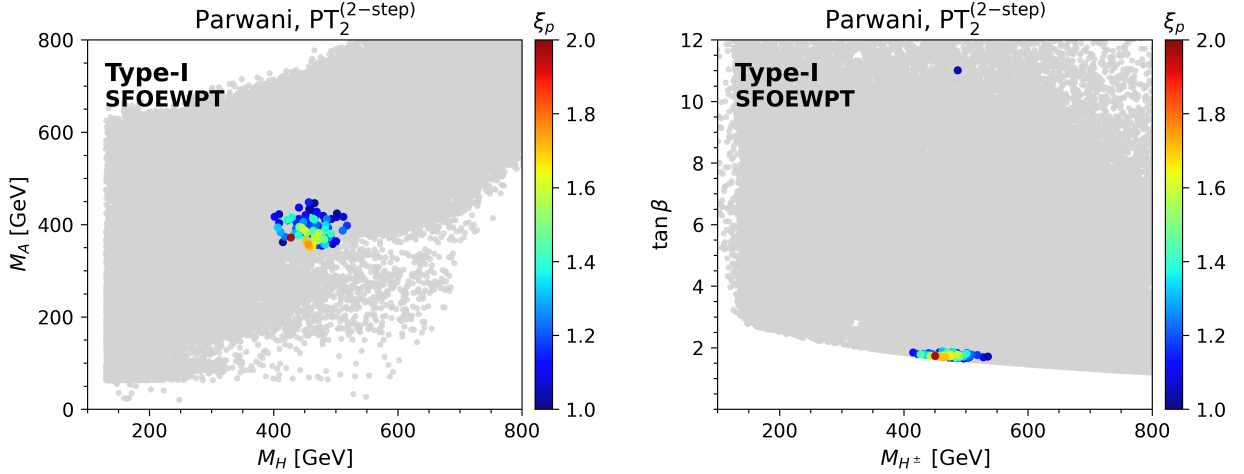}
\caption{%
Distribution of SFOEWPT parameter points for the second stage of the two-step SFOEWPT scenario ($\trans{2}{2}$) projected onto the $(M_H, M_A)$ plane (left panel) and the $(M_{H^\pm}, \tb)$ plane (right panel) for the Type-I model under the Parwani resummation scheme. The results are compiled from an expanded dataset totaling $5.4\times10^6$ baseline physical parameter points (gray points). The color scale denotes the order parameter $\xi_p$.
}
\label{fig-PT22}
\end{figure}

Finally, we turn to multi-step SFOEWPTs and investigate whether they occupy
distinctive regions of parameter space. Although two-step SFOEWPTs are
exceedingly rare in the Normal Scenario, it is nevertheless instructive to
examine their characteristic parameter-space distributions. Within the Parwani
framework, only the second stage of a two-step transition, $\trans{2}{2}$,
occurs with a meaningful event count in Type-I, yielding
$\nsfo(\trans{2}{2})=14$ from the baseline sample of $10^6$ physical
parameter points. Type-X produces an even smaller sample,
$\nsfo(\trans{2}{2})=3$, rendering a statistical analysis infeasible.
Accordingly, we focus on Type-I $\trans{2}{2}$ transitions in the following discussion.

Because the baseline sample contains too few $\trans{2}{2}$ events to
reliably infer the geometry of the underlying parameter space, we performed a
dedicated high-statistics scan, increasing the Type-I sample size to
$5.4\times10^6$ physical parameter points. From this enlarged sample, we
identified a total of 97 viable $\trans{2}{2}$ SFOEWPT parameter points. Their
distributions in the $(M_H,M_A)$ and $(M_{H^\pm},\tb)$ planes are shown in
\autoref{fig-PT22}.

The resulting $\trans{2}{2}$ parameter space is remarkably localized. The heavy
neutral and charged Higgs masses are confined to narrow intervals centered
around $M_A\simeq400\gev$ and
$M_H\simeq M_{H^\pm}\simeq470\gev$, with widths of only about
$\pm60\gev$. Likewise, $\tb$ is restricted to the narrow range
$1.6\lesssim\tb\lesssim1.95$, apart from a single outlier at
$\tb\simeq11$. Given that this isolated point appears only once among
$5.4\times10^6$ physical parameter points, the results indicate that viable
$\trans{2}{2}$ transitions are overwhelmingly concentrated within a highly
localized region of the Type-I parameter space, with only a negligible
probability of extending to substantially larger values of $\tb$.

\section{Characteristics of GW Parameter Points with the Parwani Resummation Method}
\label{sec:characteristics:GW:Parwani}

In the preceding section, we characterized the SFOEWPT parameter space in terms
of the heavy-Higgs masses, scalar mixing parameters, and the soft-breaking
parameter. However, satisfying the SFOEWPT condition alone does not guarantee
an observable stochastic GW signal, even though it ensures sufficient
suppression of sphaleron processes in the broken phase. From the perspective of
future space-based GW interferometers, we define the subset of SFOEWPT
parameter points with a four-year LISA SNR satisfying
$\mathrm{SNR}>10$ as the \textit{GW parameter points}. In this section, we
investigate the hydrodynamic properties and phenomenological characteristics of
these GW parameter points under the Parwani resummation scheme.

\subsection{Hydrodynamic Trends and Acoustic Lifetime Suppression}
\label{subsec:hydrodynamics}

\begin{figure}[t]
\centering
\includegraphics[width=0.65\textwidth]{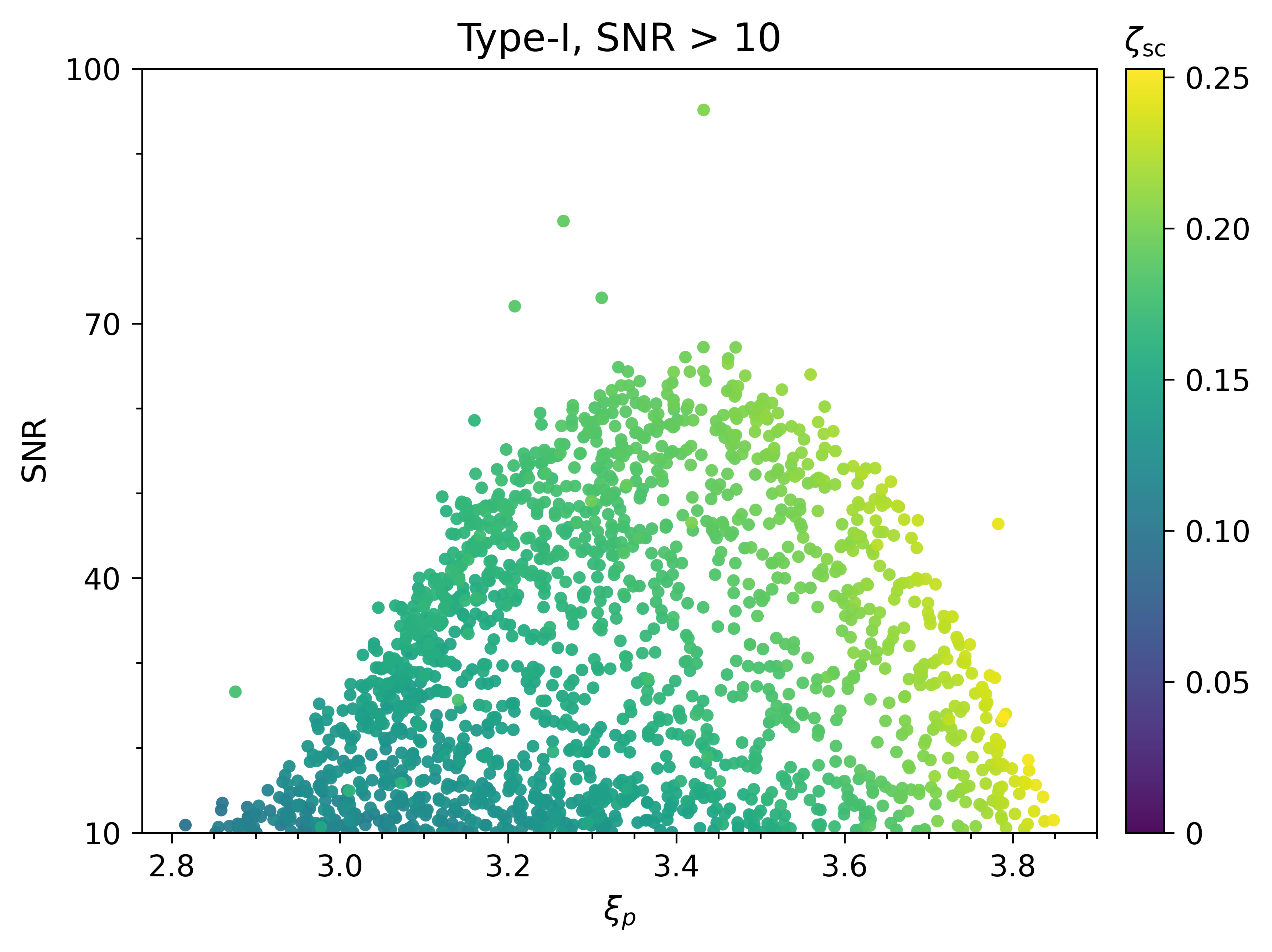}
\caption{%
Distribution of viable GW parameter points ($\text{SNR}>10$) in the $(\xi_p,\text{SNR})$ plane for the Type-I 2HDM under the Parwani resummation scheme. The color scale denotes the supercooling measure $\zeta_{\rm SC}\equiv(T_c-T_n)/T_c$.
}
\label{fig-TypeI-SNR-xip-Scooling}
\end{figure}

We begin our hydrodynamic analysis by examining the relationship between the
percolation order parameter $\xi_p$ and the resulting GW SNR.
\autoref{fig-TypeI-SNR-xip-Scooling} shows the GW parameter points in the
$(\xi_p,\mathrm{SNR})$ plane for the Type-I model under the Parwani
resummation scheme. To understand the structure of this distribution, it is
useful to consider the role of cosmic supercooling, which strongly influences
the macroscopic phase-transition parameters relevant for GW production. 
Cosmic supercooling is a generic feature of an SFOEWPT, in which the scalar
field remains trapped in the false symmetric vacuum after the critical
temperature $T_c$ is reached, until bubble nucleation eventually becomes
efficient~\cite{Kobakhidze:2017mru,Ellis:2018mja,Athron:2022mmm}. The relevant
quantity is therefore the degree of supercooling, which we quantify through the
dimensionless parameter~\cite{Athron:2022mmm}
\begin{equation}
\zeta_{\rm SC}\equiv\frac{T_c-T_n}{T_c}.
\end{equation}
The color scale in \autoref{fig-TypeI-SNR-xip-Scooling} denotes
$\zeta_{\rm SC}$.

Three characteristic features emerge from the
$(\xi_p,\mathrm{SNR})$ distribution shown in
\autoref{fig-TypeI-SNR-xip-Scooling}. First, there is no one-to-one
correspondence between $\xi_p$ and the GW SNR. Parameter points with the same
value of $\xi_p$ exhibit a wide spread in SNR, populating the region between
the observability threshold ($\mathrm{SNR}=10$) and the upper envelope of the
distribution. This behavior reflects the fact that $\xi_p$ measures only the
field-excursion ratio at percolation and therefore does not uniquely determine
the macroscopic phase-transition parameters,
$\alpha_{\rm GW}$ and $\beta_{\rm GW}/H_*$, that govern the resulting GW
spectrum.

Second, a clear positive correlation emerges between the degree of
supercooling, $\zeta_{\rm SC}$, and the order parameter $\xi_p$.
Configurations with mild supercooling
($\zeta_{\rm SC}\lesssim0.12$) are confined to the left side of the
distribution, where $\xi_p\in[2.8,3.2]$. In contrast, configurations with
stronger supercooling, reaching $\zeta_{\rm SC}\simeq0.24$, are shifted toward
larger values of $\xi_p$, extending up to $\xi_p\simeq3.86$.

The origin of this correlation can be traced to the fact that both $\xi_p$ and
$\zeta_{\rm SC}$ are governed by the structure of the potential barrier
separating the symmetric and broken phases. A larger field-excursion ratio,
$\xi_p=v(T_p)/T_p$, is generally associated with a higher and/or broader
barrier, which increases the bounce action $S_3$. Since the bubble nucleation
rate is exponentially suppressed,
$\Gamma(T)\propto\exp(-S_3/T)$,
a larger barrier delays the onset of efficient nucleation. Consequently, the
scalar field remains trapped in the false vacuum for a longer period while the
Universe continues to cool, resulting in a larger degree of supercooling before
the phase transition can successfully proceed.

Third, the observability requirement $\mathrm{SNR}>10$ imposes both lower and
upper bounds on the percolation order parameter,
$2.8\lesssim \xi_p \lesssim 3.85$. The origin of this double-sided constraint
can be understood by examining the parameter points that define the upper
envelope of the $\mathrm{SNR}$--$\xi_p$ distribution. Their behavior---namely,
that the SNR initially increases with $\xi_p$, reaches a maximum, and
subsequently decreases---gives rise to both the lower and upper bounds implied
by the detectability criterion $\mathrm{SNR}>10$. 

\begin{figure}[t]
\centering
\includegraphics[width=\textwidth]{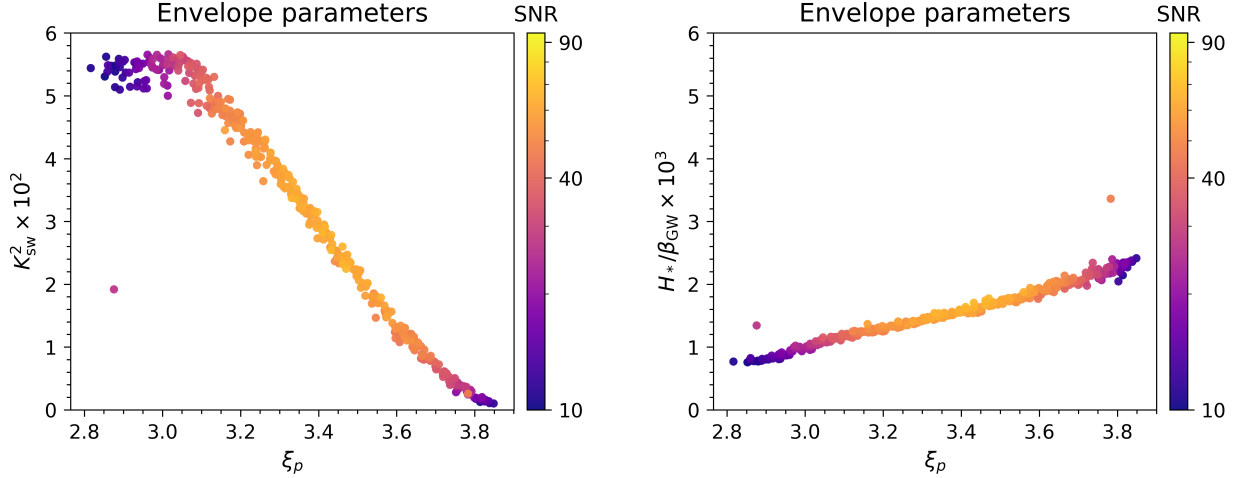}
\caption{%
Distribution of the SNR envelope parameter points projected onto the $(\xi_p, K_{\text{sw}}^2)$ plane (left panel) and the $(\xi_p, H_*/\beta_{\text{GW}})$ plane (right panel) for the Type-I 2HDM under the Parwani resummation scheme. The color scale denotes the four-year LISA $\text{SNR}$.
}
\label{fig-Type1-envelope}
\end{figure}

To understand the behavior of the envelope parameter points, we first isolate
them by dividing the accessible $\xi_p$ range into 60 equal-width bins and
selecting the parameter points whose SNR values lie within the highest 15\% of
each bin. Since the acoustic GW energy density scales approximately as
$h^2\Omega_{\rm sw}\propto K_{\rm sw}^2(H_*/\beta_{\rm GW})$,
where $K_{\rm sw}$ denotes the acoustic kinetic-energy fraction and
$H_*/\beta_{\rm GW}$ characterizes the transition duration, we project the
envelope parameter points onto the $(\xi_p,K_{\rm sw}^2)$ plane (left panel)
and the $(\xi_p,H_*/\beta_{\rm GW})$ plane (right panel) in
\autoref{fig-Type1-envelope}. The color scale indicates the corresponding SNR.
For visual clarity, the vertical axes are rescaled as
$K_{\rm sw}^2\times10^2$ and $(H_*/\beta_{\rm GW})\times10^3$.

In the lower-$\xi_p$ region, $2.8\lesssim\xi_p\lesssim3.1$, the acoustic
kinetic-energy fraction remains approximately constant, while the transition
duration increases steadily. Consequently, the envelope SNR rises with
increasing $\xi_p$. The observability requirement $\mathrm{SNR}>10$ therefore excludes parameter
points with values of $\xi_p$ below a critical threshold, establishing the
lower bound discussed above.

The behavior reverses in the stronger-transition regime, $\xi_p\gtrsim3.6$.
As $\xi_p$ increases, $H_*/\beta_{\rm GW}$ continues to grow, whereas
$K_{\rm sw}^2$ decreases. Since the suppression of $K_{\rm sw}^2$ overwhelms
the enhancement from the longer transition duration, the acoustic GW amplitude
declines and the envelope SNR decreases. As a result, the $\mathrm{SNR}>10$
criterion imposes an upper bound on $\xi_p$.

\begin{figure}[t]
\centering
\includegraphics[width=0.6\textwidth]{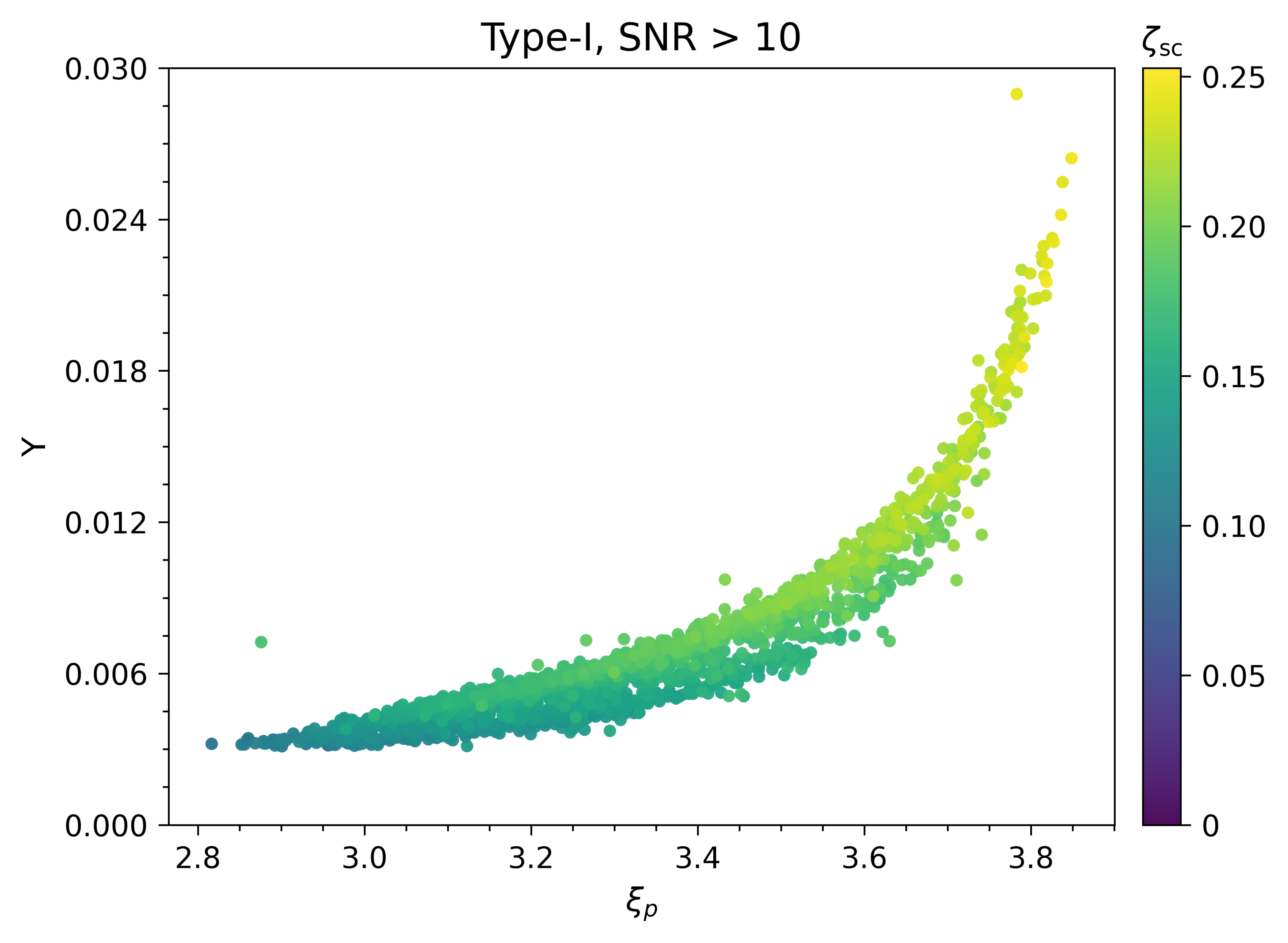}
\caption{%
The sound-wave lifetime suppression factor $\Upsilon$ as a function of the percolation order parameter $\xi_p$ across the viable GW parameter space for the Type-I 2HDM under the Parwani resummation scheme. The color scale maps the corresponding supercooling measure $\zeta_\text{SC}$.
}
\label{fig-Type1-Upsilon-xip-zetaSC}
\end{figure}

Finally, we examine the impact of the sound-wave lifetime suppression factor
$\Upsilon$, which accounts for the finite duration of the acoustic source.
\autoref{fig-Type1-Upsilon-xip-zetaSC} displays the GW parameter points in the
$(\xi_p,\Upsilon)$ plane, with the color scale indicating the supercooling
measure $\zeta_{\rm SC}$.

A striking feature of the distribution is the severity of the suppression.
Across all observable GW parameter points in the Normal Scenario of the Type-I
2HDM, $\Upsilon$ remains confined to the range
$\mathcal{O}(10^{-4})$--$\mathcal{O}(10^{-2})$. This behavior originates from
the short lifetime of the acoustic source. Early studies typically assumed that
sound waves persisted for approximately a Hubble time,
$H_*\tau_{\rm sw}\sim1$, yielding $\Upsilon\simeq1$~\cite{Hindmarsh:2013xza,Hindmarsh:2015qta,Hindmarsh:2017gnf} and therefore
maximizing the GW production efficiency. More recent hydrodynamic analyses have
shown that nonlinear steepening of fluid waves triggers shock formation and the
onset of turbulence, terminating the efficient acoustic source much earlier and
driving the system into the regime
$H_*\tau_{\rm sw}\ll1$~\cite{Ellis:2020awk,Guo:2020grp,Dahl:2024eup}.

We find that all viable GW parameter points in the Normal Scenario lie deeply
within the short-lifetime regime characterized by
$H_*\tau_{\rm sw}\ll1$.  This follows directly from the scaling of the
shock-formation timescale $\tau_{\rm sh}$ in \autoref{eq-Htaush}, which is
proportional to $H_*/\beta_{\rm GW}$ and inversely proportional to the
root-mean-square fluid velocity,
$\bar U_f\propto\sqrt{K_{\rm sw}}$. As shown in
\autoref{fig-Type1-envelope}, the GW parameter points typically exhibit
$H_*/\beta_{\rm GW}\sim\mathcal{O}(10^{-3})$ together with
$\bar U_f\sim\mathcal{O}(10^{-1})$, naturally leading to
$H_*\tau_{\rm sw}\ll1$. In this limit,
the suppression factor in \autoref{eq-Upsilon} simplifies to
$\Upsilon\simeq H_*\tau_{\rm sw}$.

A further notable feature of
\autoref{fig-Type1-Upsilon-xip-zetaSC} is the positive correlation between
$\Upsilon$ and $\xi_p$. Its origin can again be traced to the acoustic
lifetime. In the limit $H_*\tau_{\rm sw}\ll1$, the sound-wave lifetime is
controlled by the shock-formation timescale $\tau_{\rm sh}$, which scales
linearly with the transition duration $H_*/\beta_{\rm GW}$ according to
\autoref{eq-Htaush}. As shown in the right panel of
\autoref{fig-Type1-envelope}, the envelope GW parameter points exhibit a
monotonic increase of $H_*/\beta_{\rm GW}$ with increasing $\xi_p$.
Consequently, larger values of $\xi_p$ correspond to longer-lived acoustic
sources and hence larger values of $\Upsilon$. Nevertheless, even at its
maximum, the suppression factor remains of order $10^{-2}$, indicating that
the acoustic contribution to the GW signal is always strongly suppressed.

In summary, these hydrodynamic effects reveal a central feature governing the
observability of electroweak phase transitions. 
As $\xi_p$ increases, the phase transition becomes stronger and the associated
supercooling becomes more pronounced, reshaping the balance between the
transition duration and the acoustic kinetic-energy fraction. As a result, the
GW signal does not increase monotonically with $\xi_p$.
Instead,
the competition between the acoustic kinetic-energy fraction and the
transition duration causes the $\mathrm{SNR}>10$ requirement to impose both
lower and upper bounds on $\xi_p$. Furthermore, all viable GW parameter points
lie in the short-lived acoustic regime, leading to substantial sound-wave lifetime
suppression, $\Upsilon\ll1$. Consequently, within the present 2HDM scan, the GW signals accessible to future
interferometers such as LISA originate not from the strongest phase
transitions, but from an intermediate range of the percolation order
parameter, $2.8\lesssim\xi_p\lesssim3.82$,
where these competing hydrodynamic effects are optimally balanced.

\subsection{Phenomenological Implications Across the Four Yukawa Types}
\label{subsec:pheno_4types}

Having established the microphysical and hydrodynamic mechanisms that govern
the GW signal, we now translate these results into constraints on the 2HDM
parameter space. As we shall see, the regions capable of satisfying the GW
observability requirement $\mathrm{SNR}>10$ depend strongly on the Yukawa
structure of the model. In this subsection, we project the viable GW parameter
points onto the scalar-mass and mixing-parameter planes, and compare the
surviving regions across the four Yukawa types.

\begin{figure}[t]
\centering
\includegraphics[width=\textwidth]{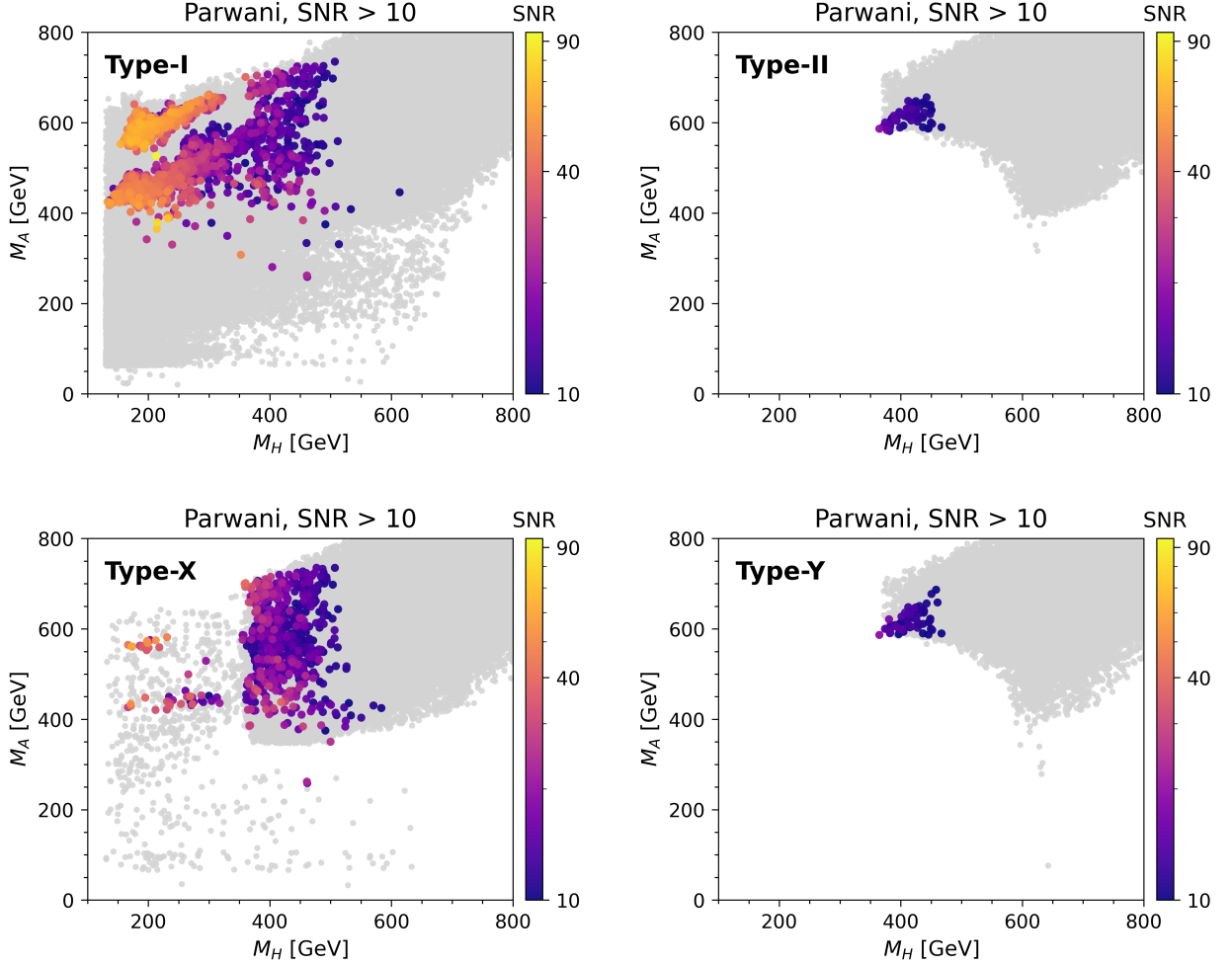}
\caption{%
Distribution of GW parameter points in the $(M_H, M_A)$ plane for the one-step transition scenario across the four 2HDM types under the Parwani resummation scheme. The color scale denotes the LISA four-year SNR. The subpanels correspond to Type-I (upper left), Type-II (upper right), Type-X (lower left), and Type-Y (lower right).
}
\label{fig-4types-MA-MH-SNR}
\end{figure}

We begin with \autoref{fig-4types-MA-MH-SNR}, which shows the viable GW
parameter points in the $(M_H,M_A)$ plane for the four Yukawa types. The color
scale denotes the four-year LISA SNR. The strongest GW signals are obtained in
Type-I, where the maximum SNR reaches 95.14. The corresponding maxima in
Type-II, Type-X, and Type-Y are 19.38, 60.41, and 19.38, respectively. Across
all four types, larger SNR values preferentially occur in the lighter-mass
region, with the dominant hierarchy satisfying $M_H<M_A$.

A second notable feature is the substantial reduction of the upper bounds on
the BSM Higgs masses in the GW parameter space relative to the corresponding
SFOEWPT region. While the SFOEWPT solutions extend up to approximately
$1.6~\tev$ in both $M_H$ and $M_A$, the requirement
$\mathrm{SNR}>10$ imposes significantly stronger upper limits. The maximal
allowed values of $M_A$ are approximately $750\gev$ in Type-I, $680\gev$ in
Type-II, $750\gev$ in Type-X, and $700\gev$ in Type-Y. Similarly, the upper
bounds on $M_H$ are approximately $500\gev$ in Type-I and Type-X, and
$460\gev$ in Type-II and Type-Y.

\begin{figure}[t]
\centering
\includegraphics[width=\textwidth]{fig-4types-tb-MHc-SNR}
\caption{%
Distribution of the viable GW parameter points in the $(M_{H^\pm}, \tb)$ plane for the one-step transition scenario across the four 2HDM types under the Parwani resummation scheme. The color scale denotes the LISA four-year $\text{SNR}$. The subpanels correspond to Type-I (upper left), Type-II (upper right), Type-X (lower left), and Type-Y (lower right).
}
\label{fig-4types-tb-MHc-SNR}
\end{figure}

Having examined the neutral-Higgs mass plane, we next turn to the charged-Higgs
sector. \autoref{fig-4types-tb-MHc-SNR} displays the viable GW parameter points
in the $(M_{H^\pm},\tan\beta)$ plane, with the color scale indicating the
corresponding SNR. As in the neutral-Higgs sector, the GW observability
requirement substantially reduces the viable parameter space relative to the
corresponding SFOEWPT region. In particular, it imposes an upper bound of
approximately $680\gev$ on the charged-Higgs mass, significantly stronger than
the $\sim1.6\tev$ limit obtained from the SFOEWPT requirement alone.

In Type-I, the surviving points form two distinct vertical bands centered at
$M_{H^\pm}\sim200\gev$ and $M_{H^\pm}\sim500\gev$. Type-X retains sparse
remnants of the same structure, reflecting the considerably smaller number of
viable GW parameter points. By contrast, Type-II and Type-Y are strongly
compressed by the combined effects of the $\mathrm{SNR}>10$ requirement and
the FCNC lower bound on $M_{H^\pm}$, leaving only the narrow interval
$M_{H^\pm}\in[600,680]\gev$.

\begin{figure}[t!]
\centering
\includegraphics[width=\textwidth]{fig-4types-tb-sba-SNR}
\caption{%
Distribution of GW parameter points in the $(s_{\beta-\alpha}, \tb)$ plane for the one-step transition scenario across the four 2HDM types under the Parwani resummation scheme. The color scale denotes the LISA four-year SNR. The subpanels correspond to Type-I (upper left), Type-II (upper right), Type-X (lower left), and Type-Y (lower right).
}
\label{fig-4types-tb-sba-SNR}
\end{figure}

Finally, \autoref{fig-4types-tb-sba-SNR} shows the viable GW parameter points
projected onto the $(s_{\beta-\alpha},\tan\beta)$ plane. The overall shapes of
the distributions closely resemble those of the corresponding SFOEWPT regions
shown in \autoref{fig-4types-tb-sba-xip}. The principal difference lies in the
allowed range of $\tan\beta$ for Types II, X, and Y. Whereas the SFOEWPT
parameter space extends to the upper limit of the scan range, the GW
requirement significantly restricts the maximum values of $\tan\beta$, yielding
approximately $\tan\beta\lesssim3$ in Type-II,
$\tan\beta\lesssim27$ in Type-X, and $\tan\beta\lesssim5$ in Type-Y.

Type-I allows the largest departure from the exact alignment limit, reaching
$|s_{\beta-\alpha}|\simeq0.96$. Within this region, however, the SNR exhibits
no significant correlation with $|s_{\beta-\alpha}|$, implying that future
Higgs-coupling measurements are unlikely to provide a useful discriminator of
the GW signal strength. By contrast, Type-II and Type-Y strongly favor the
exact alignment limit, restricting the viable parameter space to
$1-|s_{\beta-\alpha}|\lesssim\mathcal{O}(10^{-3})$.

A distinctive feature of Type-X is the persistence of the arm-like structures
already present in the SFOEWPT parameter space. These branches, originating
from the wrong-sign $\tau$-Yukawa solution, permit deviations as large as
$|s_{\beta-\alpha}|\simeq0.97$. Consequently, they can be probed through their
characteristic multi-$\tau$ signatures at the HL-LHC and future lepton
colliders~\cite{Ferreira:2014naa,Chun:2015hsa}.

\section{Conclusion}
\label{sec-conclusion}

In this work, we have performed a comprehensive comparative analysis of strong
first-order electroweak phase transitions (SFOEWPTs) and their associated
stochastic gravitational-wave (GW) backgrounds within the \textit{CP}-conserving
Two-Higgs-Doublet Model (2HDM) with a softly broken $Z_2$ symmetry under the
Normal Scenario, where the lighter \textit{CP}-even scalar is identified with
the observed 125~GeV Higgs boson. Utilizing the numerical package
\texttt{BSMPT} v3.1.8, we systematically analyzed the finite-temperature vacuum
structure, phase-transition dynamics, and stochastic GW power spectra across
all four Yukawa realizations (Types I, II, X, and Y). To construct a robust
statistical foundation, our global scan first generated a baseline sample of
$10^6$ physically viable parameter points for each Yukawa type and then
subjected them to finite-temperature analyses using two independent thermal
resummation prescriptions, namely the Parwani and Arnold--Espinosa (AE)
schemes.

To characterize the thermal and observational implications of these
configurations, we identified two nested populations of parameter points. The
first, referred to as the \emph{SFOEWPT parameter points}, satisfies the
conventional condition $\xi_p\ge1$, where $\xi_p$ denotes the ratio of the
Higgs vacuum expectation value to the plasma temperature at the percolation
epoch. The second, referred to as the \emph{GW parameter points}, consists of
the subset capable of generating an observable GW signal with an expected
four-year LISA signal-to-noise ratio satisfying $\mathrm{SNR}\ge10$.

A central outcome of our analysis arises from the comparison between the
Parwani and AE prescriptions. As a common feature, single-step transitions
overwhelmingly dominate the viable configurations under both resummation
schemes. This behavior stands in sharp contrast to the Inverted Scenario, where
multi-step transitions occur much more frequently. Beyond this common baseline,
however, a pronounced scheme dependence emerges. This theoretical uncertainty
substantially alters not only the total number of successful SFOEWPT and GW
parameter points, but also the geometric structure of the viable parameter
space. While the Parwani prescription yields continuous and well-organized
SFOEWPT regions, the AE prescription exhibits an extreme sensitivity to the
$\xi_p\ge1$ requirement, generating fragmented parameter-space distributions
and irregular voids in the heavy-scalar mass planes. These differences extend
beyond the microscopic model parameters and lead to qualitatively different
predictions for the macroscopic phase-transition parameters relevant for GW
production.

Within the Parwani framework, the SFOEWPT parameter space exhibits several
distinctive features. Most importantly, the SFOEWPT requirement strongly
restricts the conventional decoupling regime by imposing a common upper bound
on the heavy BSM Higgs spectrum,
$M_{H,A,H^\pm}\lesssim1.6~\tev$. Relative to the physically viable parameter
space, the lower mass bounds are raised in Type-I and Type-X once
$\xi_p\ge1$ is imposed, whereas they remain essentially unchanged in Type-II
and Type-Y. The SFOEWPT condition also favors the Higgs alignment limit, with a
type-dependent degree of allowed misalignment. Type-I permits the largest
deviations, of order $\mathcal{O}(10^{-2})$, while Type-II and Type-Y restrict
alignment deviations to $\mathcal{O}(10^{-3})$. Finally, two-step SFOEWPT
solutions appear only in a highly localized Type-I region, centered around
$M_A\sim400\gev$, $M_{H,H^\pm}\sim500\gev$, and $\tan\beta\sim2$.

The GW parameter space exhibits an even stronger dependence on the Yukawa
structure. Type-I produces the largest population of viable GW parameter
points, followed by Type-X. In contrast, the combined effects of the FCNC
constraints and the observability requirement $\mathrm{SNR}>10$ severely
restrict the parameter space in Type-II and Type-Y, leaving only
approximately 60 viable configurations in each case. Across all four Yukawa
types, the GW requirement imposes substantially stronger constraints on the
heavy-Higgs masses than the SFOEWPT condition alone. While the SFOEWPT
parameter space extends up to approximately $1.6~\tev$, the observable GW
parameter space is typically restricted to heavy-Higgs masses below
$\sim700~\gev$.

For the highly observable GW solutions in Type-I, we identified their characteristic hydrodynamic properties.
The acoustic source operates in the short-lifetime
sound-wave regime characterized by $H_*\tau_{\rm sw}\ll1$. Observable GW
signals arise from electroweak phase transitions with substantial plasma fluid
velocities, $\bar U_f\sim\mathcal{O}(10^{-1})$, and relatively short
transition durations,
$H_*/\beta_{\rm GW}\sim\mathcal{O}(10^{-3})$. As a consequence, the acoustic
source persists for only about $1\%$ of a Hubble time, leading to a severe
sound-wave lifetime suppression factor of
$\Upsilon\sim\mathcal{O}(10^{-2})$. Despite this suppression, a detectable GW
signal remains possible within a well-defined region of the 2HDM parameter
space.

Taken together, our results establish a direct connection, within the framework
of the 2HDM, among the finite-temperature electroweak phase transition, the
hydrodynamic evolution of the primordial plasma, and the collider-scale
parameters of the model. 
They
demonstrate that both the viability of SFOEWPTs and the resulting GW signals
depend sensitively on the treatment of thermal daisy resummation, while the
detectable GW population occupies a highly specific hydrodynamic regime
characterized by short-lived acoustic sources. Upcoming observations at LISA,
combined with direct searches for heavy Higgs bosons at future colliders, will
therefore provide powerful and complementary tests of the cosmological
viability of the 2HDM.

\acknowledgments
JC and DK are supported by National Institute for Mathematical Sciences (NIMS) grant funded by the Korea government (MSIT) (No.~B26810000).
JK is supported by a KIAS Individual Grant (PG099201) and by the Center for Advanced Computation, both at the Korea Institute for Advanced Study.
SL is supported by the National Science \& Technology Council under grant no. NSTC 113-2112-M-007-041-MY3.
\appendix

\section{Cosmological Viability Criteria within \package{BSMPT}}
\label{appendix-BSMPT-selection}

After applying the theoretical and experimental constraints described in the
main text, additional selection criteria must be imposed on the output of
\package{BSMPT} to identify cosmologically viable phase transitions. This
appendix summarizes the thermal, dynamical, and numerical requirements used in
our analysis. A parameter point is retained for the final GW analysis only if
it satisfies the following three categories of requirements:
\begin{enumerate}[label=(\roman*)]
\item The thermal history and tunneling solution should be successfully computed
without numerical failures;
\item The thermal evolution should terminate in the physical electroweak vacuum
at zero temperature;
\item A physically consistent hydrodynamic solution should exist, corresponding
to a positive sound-wave efficiency factor, $\kappa_{\rm sw}>0$.
\end{enumerate}

The first category concerns the successful numerical reconstruction of the
thermal history, which is monitored through the internal status flags provided
by \package{BSMPT}. We require all retained parameter points to satisfy the
following conditions:

\vspace{0.7em}
\noindent
\begin{tabularx}{\textwidth}{XX}
\texttt{status\_nlo\_stability} $==$ \texttt{success}, &
\texttt{status\_ewsr} $==$ \texttt{ew\_sym\_res}, \\
\texttt{status\_tracing} $==$ \texttt{success}, &
\texttt{status\_coex\_pairs} $==$ \texttt{success}, \\
\texttt{status\_crit\_i} $==$ \texttt{success}, &
\texttt{status\_bounce\_sol\_i} $==$ \texttt{success}, \\
\texttt{status\_nucl\_i} $==$ \texttt{success}, &
\texttt{status\_perc\_i} $==$ \texttt{success}, \\
\texttt{status\_compl\_i} $==$ \texttt{success}, &
\texttt{status\_gw\_i} $==$ \texttt{success},
\end{tabularx}
\vspace{0.5em}

\noindent where the subscript \texttt{i} denotes the target stage of the phase
transition. For the precise definitions and technical conditions associated
with these status flags, we refer the reader to the \package{BSMPTv3}
manual~\cite{Basler:2024aaf}.

Two status flags deserve additional comment. Since \package{BSMPT} version
3.1.5, the condition
\texttt{status\_gw\_i$\,==\,$success}
guarantees a positive latent-heat release,
$\alpha_{\rm GW}\ge0$. In addition, this flag excludes extremely slow
transitions with $\beta_{\rm GW}/H_*<1$, corresponding to phase transitions
whose characteristic duration exceeds a Hubble time.

The second category excludes parameter points whose thermal evolution does not
terminate in the physical electroweak vacuum. In extended scalar sectors such
as the 2HDM, the finite-temperature evolution can terminate in a vacuum
different from the desired zero-temperature electroweak minimum. To eliminate
such cases, we require the zero-temperature vacuum expectation value obtained
after thermal evolution to reproduce the physical tree-level input,
\begin{equation}
\big| v - v^{\mathrm{tree}} \big|_{T=0} < 1~\gev.
\end{equation}
Parameter points that violate this condition are discarded.

The final category requires a strictly positive sound-wave efficiency factor,
$\kappa_{\rm sw}>0$. In \package{BSMPT}, $\kappa_{\rm sw}$ is obtained from a
hydrodynamic analysis that depends on both the transition strength
$\alpha_{\rm GW}$ and the assumed bubble-wall velocity. During our scan, some
parameter points return $\kappa_{\rm sw}=0$. This indicates that the
hydrodynamic matching conditions do not admit a consistent steady-state fluid
solution for the corresponding transition within the model-independent
energy-budget framework implemented in
\package{BSMPT}~\cite{Giese:2020znk}. We therefore discard all parameter points
with $\kappa_{\rm sw}=0$, ensuring that every retained GW parameter point
admits a physically consistent hydrodynamic solution.


\end{document}